\documentstyle [12pt,psfig] {article}
\renewcommand{\baselinestretch}{1}
\newcommand{\ltsimeq}{\raisebox{-0.6ex}{$\,\stackrel
{\raisebox{-.2ex}{$\textstyle <$}}{\sim}\,$}}
\newcommand{\gtsimeq}{\raisebox{-0.6ex}{$\,\stackrel
{\raisebox{-.2ex}{$\textstyle >$}}{\sim}\,$}}

\begin{document}

\begin{center}
\Large{\textbf{UV circular polarisation in star formation regions: 
the origin of homochirality?}}
\end{center}

\begin{center}
\large{P.W.Lucas$^1$, J.H.Hough$^1$, Jeremy Bailey$^2$, 
Antonio Chrysostomou$^1$}
T.M.Gledhill$^1$, Alan McCall$^1$ 
\end{center}

\small
(1) Department of Physical Sciences, University of Hertfordshire, College Lane,
Hatfield AL10 9AB, UK.\\
(2) Anglo-Australian Observatory, Post Office Box 296, Epping, New South Wales 121,
Australia\\
\normalsize

\vspace{4mm}
\textbf{Ultraviolet circularly polarised light has been suggested as
the initial cause of the homochirality of organic molecules in terrestrial
organisms, via enantiomeric selection of prebiotic molecules 
by asymmetric photolysis. We present a theoretical investigation of 
mechanisms by which ultraviolet circular polarisation may be 
produced in star formation regions. In the scenarios considered
here, light scattering produces only a small percentage of net circular
polarisation at any point in space, due to the forward throwing nature
of the phase function in the ultraviolet. By contrast, dichroic 
extinction can produce a fairly high percentage of net
circular polarisation ($\sim 10\%$) and may therefore play a key role 
in producing an enantiomeric excess.}

\section{Introduction}

The homochirality of amino acids and sugars in terrestrial organisms
is one of the more enduring mysteries concerning the origin of life.
The sugars in DNA are exclusively D-sugars and structural proteins and
enzymes contain only L-amino acids. Far from being a mere curiosity, 
homochirality appears to be a necessity for biochemical reactions to occur 
efficiently and may well be a prerequisite for life (Joyce et al. 1984). While some 
believe that homochirality may have arisen from an initially
racemic mixture on the primordial Earth, the discovery of enantiomeric 
excesses of L-amino 
acids in the Murchison (Cronin \& Pizzarello 1997; Engel \& Macko 1997) and 
Murray (Pizzarello \& Cronin 2000) meteorites argues for an 
extraterrestrial mechanism (see Bonner 1991).
These excesses range from several per cent up to 15\%. They are not now 
thought to have been produced directly by asymmetric photolysis (Pizzarello
2002, private comm.), since the circular dichroism of the amino acids is 
probably too small to produce such a large 
excess without dissociating $>99\%$ of the molecules. Nevertheless it 
remains very possible that asymmetric photolysis introduces a smaller 
enantiomeric excess, which is then amplified by another physical or chemical
mechanism.

In our view the most promising extraterrestrial mechanism is the
asymmetric photolysis of prebiotic molecules by circularly polarised
UV light in dusty  nebulae around Young Stellar Objects (YSOs). This
mechanism was suggested by  Bailey et al.(1998)
following the observation of nebulosity with 17\% infrared circular
polarisation (CP) in a dense cluster of YSOs in the well studied BN-KL 
region of the Orion Nebula BN-KL object and close to the highly
obscured source IRc2. The crowded nature of the young star cluster in
the vicinity of the circularly polarised nebulosity (designated BNKL
SEBN) in Orion provides a  ready supply of young solar systems which
could be undergoing asymmetric photolysis at the present time. Almost 
20\% CP has since been observed in another
YSO, NGC6334V (Menard et al., 2000) but most other  YSOs for which
spatially resolved measurements exist exhibit somewhat lower  CP ($\le
5\%$). At present it appears that infrared CP ranges from 0\% to about
20\% in YSOs, though we caution that the observational database is
still very  small and the frequency of the higher value is unknown. We
note that other possible sources of CP have been considered, eg. 
synchrotron or gyro-synchrotron radiation from a supernova remnant surrounding 
a pulsar (Rubenstein et al., 1983). However observed optical CP from the Crab
Nebula is $<1\%$, which is unsurprising given the difficulties of producing
significant {\it net} CP in supernova remnants (Roberts 1984).

Unfortunately, the observations have to be made at infrared
wavelengths because of the high extinction by the dusty medium in star
formation regions. In this paper we calculate the net percentage of
CP at ultraviolet wavelengths which can occur at points in the vicinity
of YSOs and interstellar nebulae, considering a variety of scenarios.

The analysis of the Orion data by Bailey et al.(1998) and Chrysostomou et al.(2000)
considered only light scattering as the mechanism to produce CP, 
focussing on the high CP which can be produced by scattering from aligned
non-spherical dust gains. This was a natural assumption based on the 
observational data (see Section 4).
However, CP can also be produced by dichroic extinction of linearly
polarised light. This has been observed toward astrophysical nebulae
(Martin et al., 1972) and stars (eg. Kemp \& Wolstencroft 1972) and was
attributed to dichroic extinction by a twisting magnetic field. 
CP attributed to this mechanism has also been observed towards YSOs,
including the BN object in Orion. This source has CP of 1.6\%, 
measured by Lonsdale et al.(1980) in a 10 arcsec beam. More recent spatially 
resolved observations (see Bailey et al., 1998; Chrysostomou et al., 2000) support this 
interpretation since the highly polarised nebulosity in the region is 
not close enough to the BN point source to contaminate the Lonsdale 
measurement with scattered light. 

However, it remains unclear whether the CP in the nebulae is caused 
by dichroic scattering or by dichroic extinction of light which has been 
linearly polarised by scattering, a mechanism which can produce CP even if the 
magnetic field is uniform.

Dichroic extinction plays an important role in the 
calculations shown here, incorporating both multiple scattering and 
dichroic extinction. It has previously been modelled by Martin (1974) (not
including scattered light) and at infrared wavelengths by Whitney \& Wolff 
(2002) and Lucas (2003).

When the light from all parts of a system is summed, there can only be a net 
polarisation if there is some structural asymmetry. In many physical systems 
such asymmetries are small or average out over time. Brack (1998) noted that 
on Earth, net asymmetric photolysis and other electromagnetic interactions 
``probably cancel their effects on a time and space average.'' It is therefore 
easy to produce models 
which lead to negligible net CP. In this paper we search for scenarios which 
can produce significant net CP, subject to the proviso that these appear to 
be realistic and consistent with the present knowledge of star formation 
regions.

\section{Dichroic Scattering and Dichroic Extinction}

In this work we consider two basic physical mechanisms which can 
generate a high degree of CP in star forming regions: scattering by aligned 
non-spherical grains (hereafter dichroic scattering) and dichroic
extinction of linearly polarised light. Both mechanisms require 
the presence of non spherical grains which are at least partially 
aligned with respect to a common direction. The mechanism for initial
alignment of the grains is a subject of debate (eg. Purcell 1979; Lazarian 
1995, 2000; Roberge 1995) 
but it is generally accepted that non-spherical grains become aligned 
such that their axis of greatest rotational angular momemtum precesses about 
the direction of the ambient magnetic field. There is abundant observational
evidence for aligned grains in star forming regions, which produce linear 
polarisation through dichroic extinction or by emission at wavelengths from 
the ultraviolet through to the sub-millimeter (eg. Hough \& Aitken 2003).

We note that scattering or absorption by optically active materials in the 
dust grains might also generate significant CP (if they are sufficiently 
abundant) but this is not considered here.

Before embarking on more complicated multiple scattering and
extinction calculations (see Section 3) it is useful to provide an
overview of CP  production by individual spheroidal grains, which are
the simplest  non-spherical shape (see also Gledhill \& McCall 2000).
To calculate the (2$\times$2 amplitude matrix, 4$\times$4 Stokes
matrix and  4$\times$4 extinction matrix) the ampld.f and amplq.f
codes of Mishchenko (see http://www.giss.nasa.gov/$^{\sim}$crmim; Mishchenko,
Hovenier \& Travis 2000)  were employed and to a lesser extent the
codes of  Barber \& Hill (1990). Both works employ the T-matrix
approach, expanding the  electromagnetic wave as a series of spherical
harmonics.  Calculations were performed for grain axis ratios up to
3:1, considering a  wide range of size parameters and arbitrary
imaginary component of refractive index.

\subsection{Dichroic Scattering}

The polarisation state of the photon is described by the (I,Q,U,V) Stokes 
vector, defined relative to the scattering plane. It is modified by 
scattering as follows: 

\hspace{1cm} $\left(\begin{array}{c}I\\Q\\U\\V \end{array}\right)_{scat} = 
{\bf Z} \hspace{2mm}
\left(\begin{array}{c}I\\Q\\U\\V \end{array}\right)_{inc} $

\vspace{4mm}
where the Stokes matrix ${\bf Z} = \left(\begin{array}{cccc}
Z_{11} & Z_{12} & Z_{13} & Z_{14}\\ 
Z_{21} & Z_{22} & Z_{23} & Z_{24}\\
Z_{31} & Z_{32} & Z_{33} & Z_{34}\\
Z_{41} & Z_{42} & Z_{43} & Z_{44}
\end{array}\right)$

The elements of {\bf Z} are all functions of the scattering angles and
grain orientation angles There are 3 independent angles for the oblate 
spheroids used in our calculation: photon polar deflection angle, $D$; 
grain polar orientation, $\beta$; and grain azimuthal orientation ($\alpha$),
which is defined relative to the azimuthal photon deflection angle.
$\beta=90^{\circ}$ presents the most elliptical cross section to the incident
photon in the case of oblate spheroids, while $\beta=0^{\circ}$ corresponds
to a circular cross section. Full details of the definitions are given by Lucas 
(2003).

The ratio V/I gives the degree of circular polarisation,
with positive V representing left-handed polarisation (LCP), and negative V 
representing right-handed polarisation (RCP), for a thumb pointing in the 
photon's direction of flight (i.e. along the Poynting vector).\footnote{The sense of 
positive V was erroneously stated as right-handed in Lucas (2003), but plots of 
positive and negative V in that work remain correct.} Contributions
to a Stokes vector from different sources of radiation are linearly additive, 
so equal quantities of LCP and RCP simply cancel out.

A photon which is scattered by a non-spherical dust grain can gain high CP
(defined as $>10\%$)
owing to the phase difference between contributions to the electromagnetic
field originating from different parts of the grain. By contrast, spherical 
grains can produce CP only from multiple scattering and even then usually
at a low level (eg. Shafter \& Jura 1980; Fischer, Henning \& Yorke 1994; 
Lucas \& Roche 1998). The degree of CP 
depends upon the direction of scattering and the orientation of the grain.
In general, however, the greater the departure from spherical shape, the 
higher the CP (see Gledhill \& McCall 2000).
For simple grain shapes like spheroids and ellipsoids, small deflection 
angles always produce low CP, as well as low LP. The
size parameter, $x$, is defined by $x=2 \pi a/\lambda$, where $\lambda$ if
the wavelength and $a$ is the grain dimension (defined here as the radius 
of the equivalent surface area sphere).

\begin{figure*}[hp!]
\begin{center}
\begin{picture}(200,200)

\put(0,0){\includegraphics{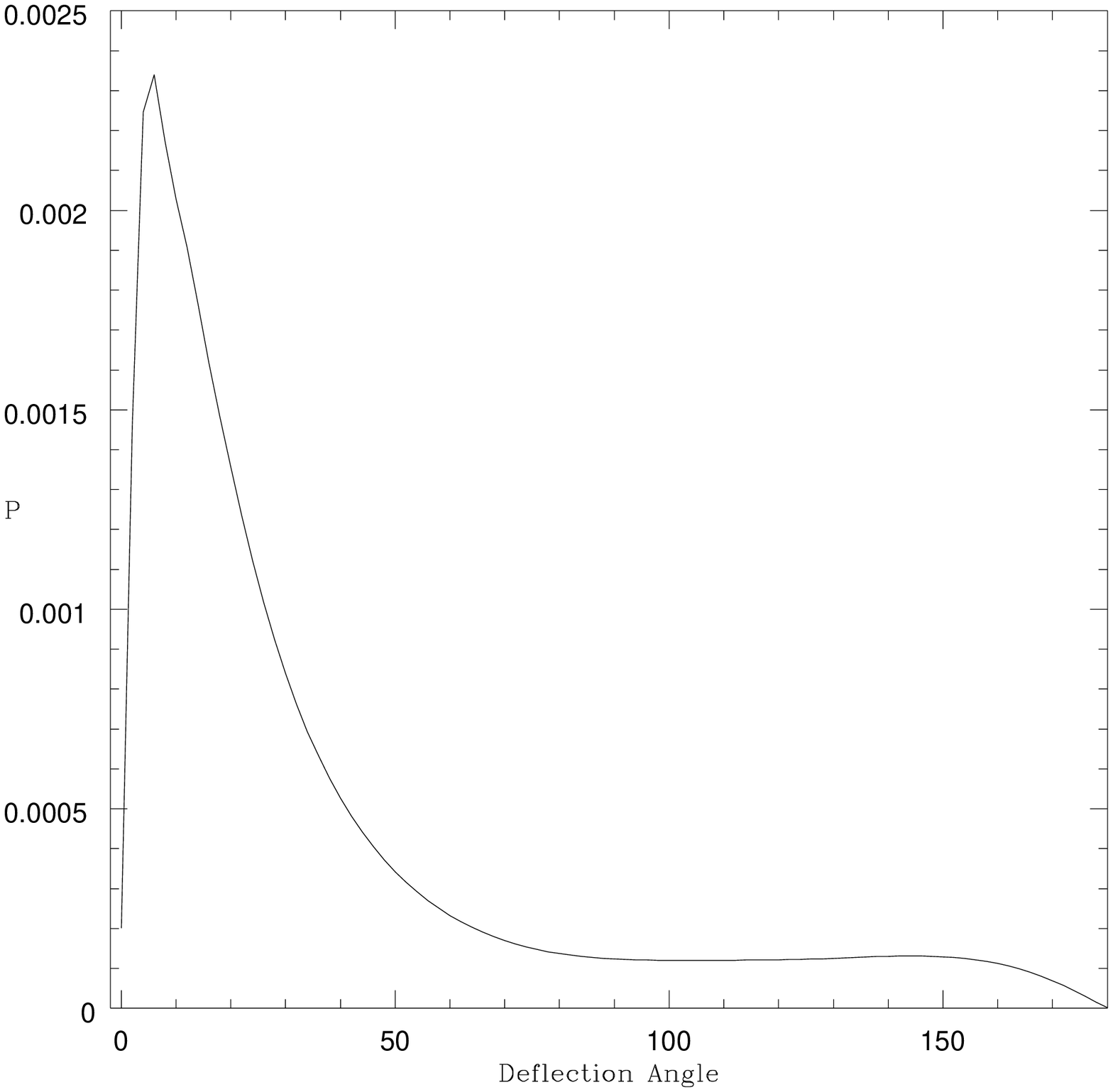}}

\put(0,0){\includegraphics{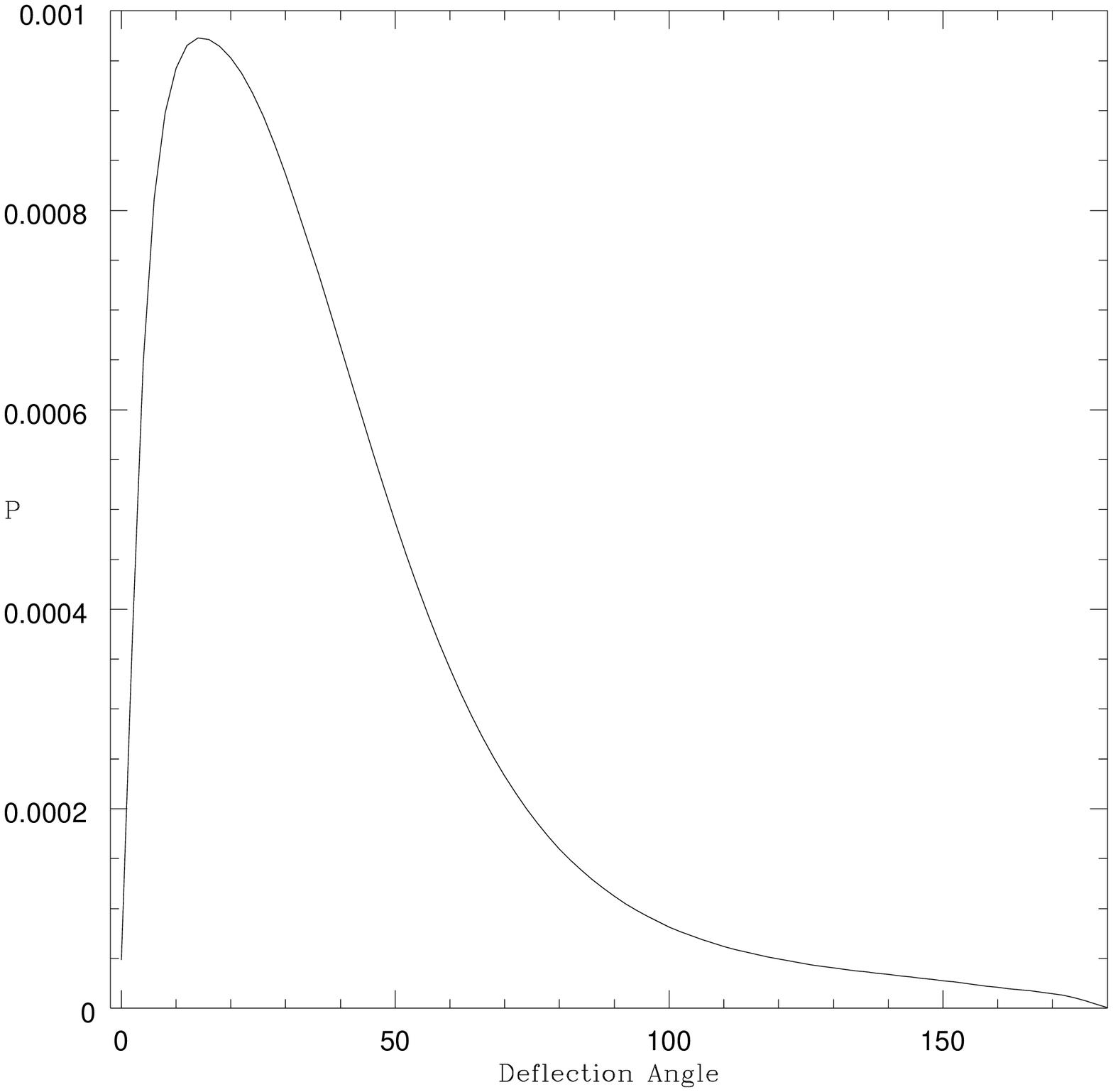}}

\end{picture} 
\end{center}
\small Figure 1. Phase function for dust in the envelope of a YSO,
showing the relative probability of scattering through different
deflection angles. The calculation uses the SACS2 mix (see section 3.1) and an 
unpolarised beam at $\lambda=0.22~\mu$m for oblate spheroids with a 3:1 axis ratio.
The plotted functions are averaged over all azimuthal scattering angles and weighted 
by the available solid angle at each deflection angle, which is proportional to 
sin$(D)$. (left) plot for face on orientation to the incident ray ($\beta=0^{\circ}$),
illustrating the highly forward throwing phase function.  
(right) edge on orientation ($\beta=90^{\circ}$), showing slightly weaker forward 
scattering due to the smaller cross section. 

\end{figure*}
Dust grains which are significantly larger than the photon wavelength
($x>1$) have a high probability of causing small angle deflection, 
producing low CP (see Figure 1). For scattering at the typical wavelengths of 
chiral bands
in prebiotic molecules (150 to 220~nm) this is likely to be the situation for 
dust grains in star formation regions, where the maximum grain size is larger 
than in the diffuse interstellar medium, perhaps reaching 0.8~$\mu$m in 
equivalent radius in the circumstellar envelopes of YSOs 
(Minchin et al., 1991a; 1991b). We note that if a photon is deflected through a large 
angle, CP close to 100\% can in principle be produced by large grains 
($x\ge1.5$ and axis ratio 2:1, see Gledhill \& McCall 2000). However, this is unlikely to 
occur in astrophysical media where there is usually a range of grain sizes. The reason is 
that grains of different size produce maximum CP at different grain orientations and 
scattering angles, so that the CP at any particular scattering angle is usually well below 
50\% (see section 3.1).

At very small size parameters ($x \ll 1$), appropriate to  very small
dust grains, the distribution of scattered light (phase function) is
more isotropic but the CP is low ($<10\%$) for all scattering angles (see
Figure 8 of Gledhill \& McCall 2000) unless the grains are very absorptive. For small 
and highly absorptive grains there is little scattered light due to the very low 
albedo. If dichroic scattering is to produce efficient asymmetric photolysis in a star 
forming region both high CP and a non-negligible albedo are desirable, so that the 
scattered light dominates the radiation field. This is most likely to occur if the
optically dominant grain size is comparable to or slightly smaller
than the wavelength, i.e. $x \sim 1$.

\subsection{Dichroic Extinction}

The polarisation state of a photon (this time defined in the reference frame of 
the grains, with positive Stokes Q parallel to the axis of grain alignment) is 
modified by dichroic extinction  when travelling a distance $s$ through a uniform
cloud with grain number density  $n$ as follows (see Whitney \& Wolff
2002; Martin 1974):

\vspace{4mm}
(2) $I_{ext} = 0.5\{(I_{inc} + Q_{inc})exp[-n (K_{11} + K_{12}) s] + (I_{inc} - Q_{inc})exp[-n (K_{11} - K_{12}) s]\}$

\vspace{4mm}
(3) $Q_{ext} = 0.5\{(I_{inc} + Q_{inc})exp[-n (K_{11} + K_{12}) s] - (I_{inc} - Q_{inc})exp[-n (K_{11} - K_{12}) s]\}$

\vspace{4mm}
(4) $U_{ext} = exp[-n K_{11} s].\{U_{inc} cos(n K_{34} s) - V_{inc} sin(n K_{34} s)\}$ 

\vspace{4mm}
(5) $V_{ext} = exp[-n K_{11} s].\{V_{inc} cos(n K_{34} s) + U_{inc} sin(n K_{34} s)\}$ 

\vspace{4mm} where $K_{11}$, $K_{12}$ and $K_{34}$ are the 3
independent elements of the  4~$\times$~4 extinction matrix which acts
on the Stokes vector. For spheroids  the $K_{ij}$ elements are
function of the grain azimuthal and polar orientation  angles
($\alpha$ and $\beta$ respectively, see Mishchenko et al., 2000.)

For unpolarised light eqs.(2-5) merely lead to light which is linearly
polarised in Stokes Q, parallel to the axis of grain alignment. Equs. (4-5)
show that CP is produced only by conversion of Stokes U to Stokes V
due to the birefringence term $K_{34}$. This term acts by introducing
a phase difference between the components of the electromagnetic wave
oscillating parallel to and perpendicular to the axis of grain
alignment. Hence, dichroic extinction only leads to CP if there is
prior linear polarisation in Stokes U. Two obvious ways of producing
this are by scattering or by prior dichroic extinction with a
different grain alignment axis. The latter case would occur if light
is passed through a nebula with a twisting magnetic field (Martin
1974) or by passing light through two spatially separated nebulae with
different magnetic field directions.

CP production by dichroic extinction is most efficient at wavelengths
significantly larger than the grain size ($x<1$), where the ratio
$K_{34}/K_{11}$ is high (see Figure 2). This initially appeared discouraging 
for the production of high CP at $\lambda \le 0.22~\mu$m in star formation
regions. We note, however, that high CP  can be produced even at UV 
wavelengths where $K_{34}/K_{11}$ is low,  provided that the ratio of
birefringence to linear dichroism ($K_{34}/K_{12}$  terms) is not far
below unity and the optical depth of the medium is  greater than unity.

\begin{figure*}[thbp]
\begin{center}
\begin{picture}(200,250)

\put(0,0){\includegraphics{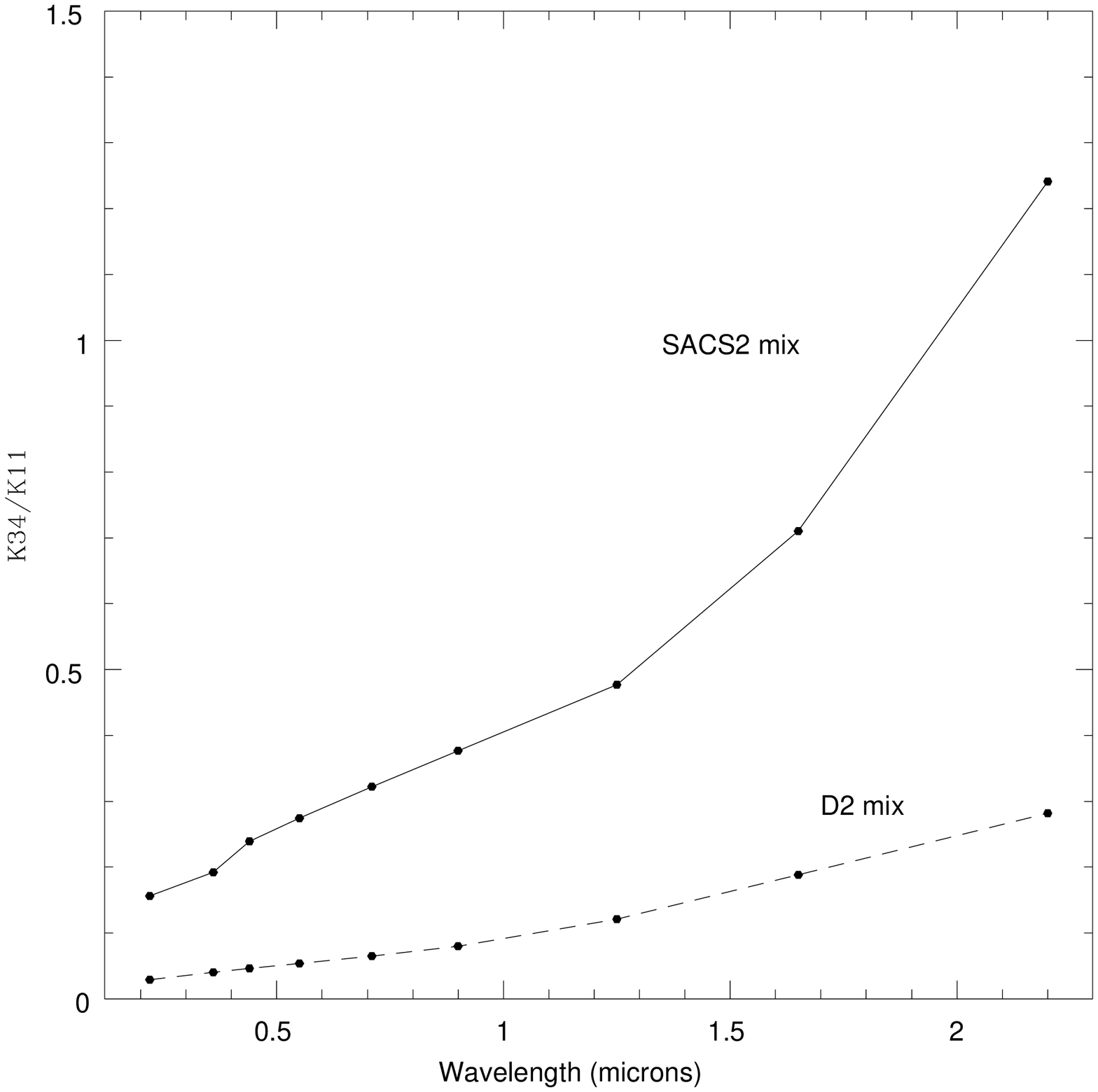}}

\put(0,0){\includegraphics{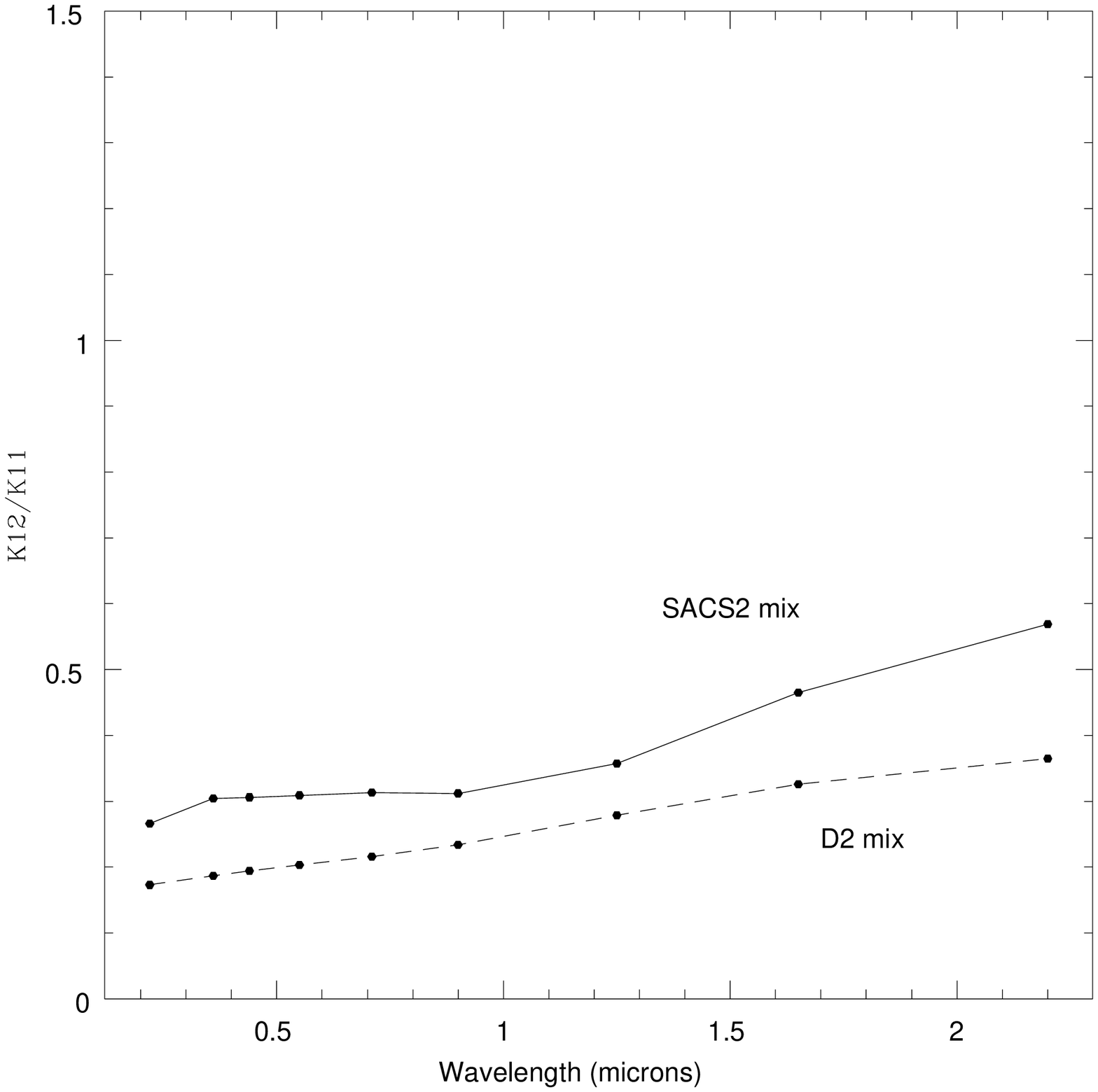}}

\end{picture} 
\end{center}
\small
\vspace{-1.5cm} Figure 2. (left) Plot of $K_{34}/K_{11}$
(birefringence/neutral extinction) as a function of wavelength. The
calculations are for the larger grain mixtures found in the
circumstellar envelopes of YSOs (see section 3.1 for definitions of grain mixtures), 
for oblate spheroids with a 3:1 axis ratio. Birefringence is weak in the ultraviolet. 
However, it is stronger for the 'shiny' SACS2 mix than the 'dirty' D2 mix. (right)
Plot of $K_{12}/K_{11}$ (dichroic extinction/neutral
extinction). Both $K_{34}/K_{11}$, $K_{12}/K_{11}$ are calculated in
the reference frame defined by the grain alignment (see text).
\end{figure*}

This point is illustrated in Figure 3 for a 220~nm light ray which is initially
100\% linearly polarised in Stokes U. It shows the change in the Stokes
vector as it passes deep into an optically thick cloud of oblate spheroid 
grains with axis ratio 3:1, which are perfectly aligned with the Stokes Q 
direction and oriented edge on to the incident photon. The chosen grain 
mixture is composed of silicate and small amorphous carbon grains 
(the SACS1 mix, see Section 3) whose size distribution is plausible 
for the interstellar environment of star forming dark clouds. For this 
mix $K_{34}/K_{11}=0.181$, while $K_{12}/K_{11}=0.294$ (evaluated at 
$\alpha=0^{\circ}$, $\beta=90^{\circ}$).

As optical depth, $\tau$, increases from zero, the K$_{34}$ term gradually
converts Stokes U to Stokes V, while the K$_{12}$ term converts U to Q
at a similar rate. Since $K_{12}/K_{11}$ is small, a high CP can be generated
at $1 < \tau < 7$, for it is not until $\tau>7$ that essentially all the
polarisation is forced into the Stokes Q plane, where oscillating electric 
waves suffer least extinction. In this example a CP of 37.5\% occurs at 
$\tau=3$, at which point 5\% of the original flux remains in the beam. For 
grain mixtures with very low absorptivity 
and similar size distribution the maximum CP approaches 50\%. However,
this calculation, which is similar to those of Martin (1974), does not 
include the diluting effect of scattered light (see Section 3.3).

\pagebreak

\begin{figure*}[thbp]
\begin{center}
\begin{picture}(200,220)

\put(0,0){\includegraphics{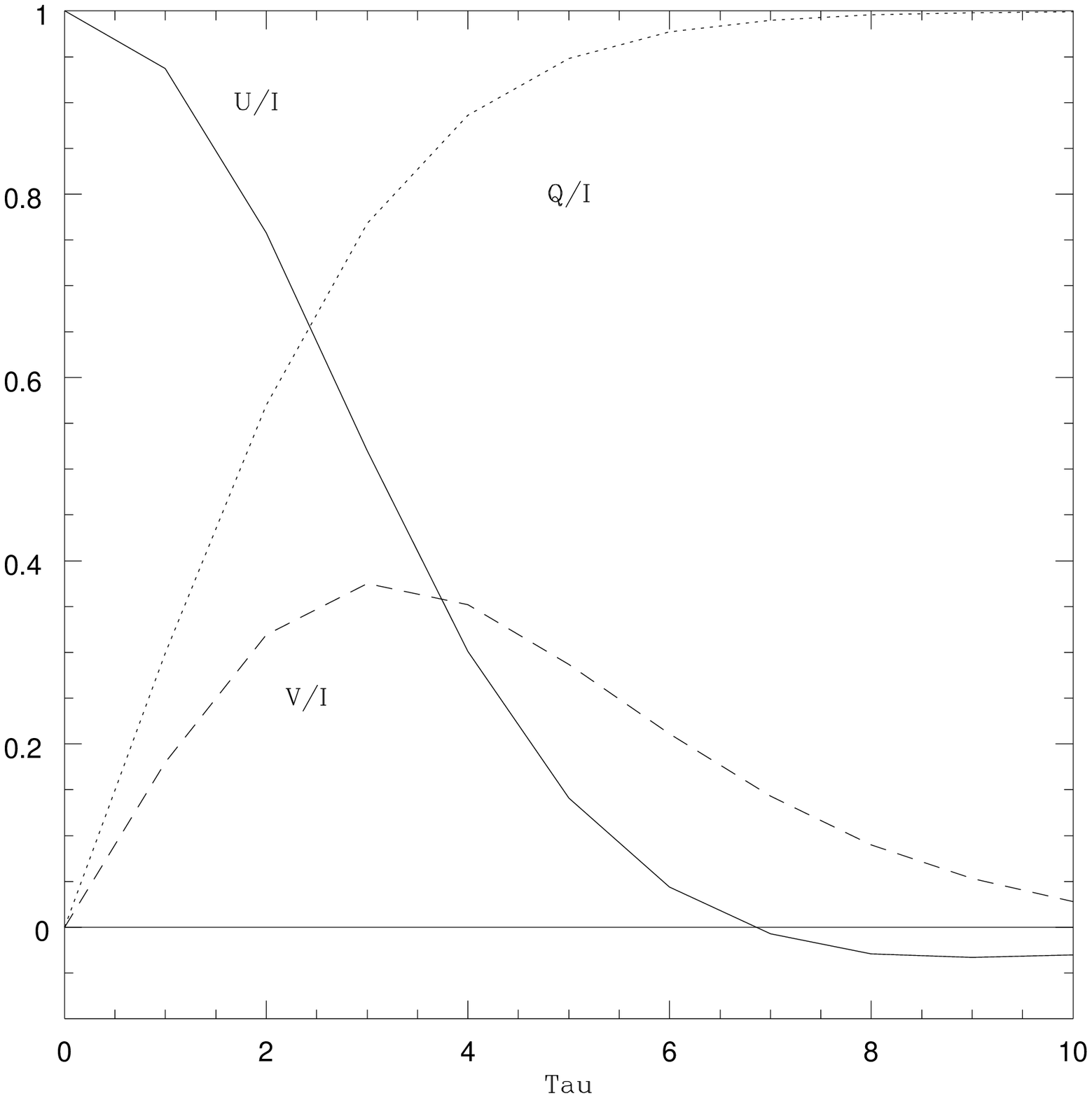}}

\end{picture} 
\end{center}
\small
\vspace{-1.5cm} Figure 3. Circular Polarisation from Dichroic Extinction.
The example calculation is at 0.22~$\mu$m, showing the effect of small, 
shiny grains (SACS1 mix) as a function of optical depth on a ray which is initially 
100\% linearly polarised (see text). The incident ray is polarised in the U plane and the 
grains are aligned with the Q plane. (solid curve) Stokes U/I; (dotted curve) Q/I; 
(dashed curve) V/I. 
\end{figure*}

\normalsize
\section{Monte Carlo modelling}



We employ the Monte Carlo method to simulate multiple scattering and
extinction in YSOs and in interstellar nebulae. The 2-D scattering 
code used here for axisymmetric nebulae is described in Lucas (2003). 
Efficient 3-D models for less symmetric nebulae have been 
implemented for YSO modelling using forced scattering (Wood \& Reynolds
1999; Lucas et al., 2003) but here we treat asymmetric
nebulae in a very simple way, which is described later.
In summary, the code randomly generates millions of photons at the
stellar surface (assuming uniform surface brightness) and each follows a 
random path through the nebula,
perhaps scattering several times before it escapes the system or is
absorbed. At present the code handles only perfectly aligned oblate
spheroids. Perfectly aligned oblate spheroids appear to yield the highest 
degrees of CP (Gledhill \& McCall 2000; Whitney \& Wolff 2002), since the 
shape presented to the wavefront does not change with rotation. Rotation 
(in the case of prolate spheroids) and precession both change the shape 
presented to the wavefront, leading to a reduction in the CP produced
after averaging over a full period.

Arbitrary magnetic field structures can be included, these being specified 
algebraically. The {\bf Z} and {\bf K} matrix elements and the amplitude 
matrix elements needed in the phase function are stored as large look up 
tables, sampling all the relevant angles every 2$^{\circ}$. This interval is 
the maximum for simulations in the ultraviolet, where the phase function
is strongly peaked in the forward scattering direction (see Figure 1).

The millions of photons output from the system are used to construct
polarimetric images of the nebula as seen in the far field, for comparison
with telescopic observation from Earth. We also calculate the net CP 
received at points close to the nebula in the near field, summing the 
contributions to the radiation field of LCP and RCP that are incident from 
all directions. Note that we have not extended the calculations to wavelengths
shorter than 0.22~$\mu$m (appropriate for the longer wavelength chiral bands) 
because: (a) computation for the shorter wavelength
chiral bands near 0.15~$\mu$m is very demanding in several respects and; (b) 
the luminous flux from all but very hot stars declines steeply at shorter 
wavelengths (see Bailey et al., 1998).

\subsection{Dust Mixtures}

The size, shape and composition of dust in star formation regions is
not known with precision and may well vary in different environments.
In this section we define four dust grain mixtures which span most of the plausible
range of size and of absorptivity that could be enountered. The two smaller mixtures
are better suited to production of CP by dichroic extinction than their larger 
counterparts. The two larger grain mixtures have a size distribution specifically 
designed to maximise CP production by dichroic scattering. 

There is a large literature on the subject which assists in constructing 
plausible dust models. Pollack (1994) gives a cogent analysis of the possible grain 
compositions in cold molecular clouds and grain sizes have been analysed by Kim, 
Martin \& Hendry (1994), among others. The size distribution in the diffuse 
interstellar medium appears to be
well modelled grains with sizes $0.005 < a < 0.25~\mu$m and a power law index 
k=3.5 in that range, so that $n(a) \propto a^{-k}$ (Mathis, Rumpl \& Nordsieck
1977). Interstellar grains in dark clouds are known to be
slightly larger than in the diffuse interstellar medium (eg. Whittet 1992): the ratio of 
total to selective extinction at visible wavelengths (R$_{V}$) rises from
3.1 in the diffuse medium to between 4 and 5.5 in different dark clouds; 
also the degree of LP due to extinction, described by the 
Serkowski law, peaks at a slightly longer wavelength in dark clouds. However,
Cardelli, Clayton \& Mathis (1989) showed that the wavelength dependence of
{\it infrared} extinction in dark clouds is similar to that of the diffuse
medium, strongly implying that there are few grains with $a > 1~\mu$m.
For interstellar dust in dark clouds we adopt the following basic size 
distribution: $0.005 < a < 0.3~\mu$m, retaining the k=3.5 power law index of 
Mathis, Rumpl \& Nordsieck. An axis ratio of 3:1 is adopted for these oblate
spheroids (see below).

We define two different mixtures of these relatively small grains for use in the 
medium of dark clouds. 

$\bullet$ the D1 mix: a highly absorptive mixture of unspecified
composition with  refractive index n=1.5-0.2i.

$\bullet$ the SACS1 mix: a more reflective mixture consisting mostly
of 'astronomical silicate'  with $0.005 < a < 0.3~\mu$m (refractive index at each
wavelength from Draine 1985) and an additional component of small
amorphous carbon grains (refractive indices from Preibisch et
al., 1993), only in the size range  $0.005 < a < 0.03~\mu$m. This grain
composition is not intended to be realistic but merely representative of
mixtures with relatively low absorption. The small carbon grains
(whose size is limited by hydrogenation and chemical sputtering in dense 
clouds, see Sorrell (1990)) have very little influence on the
polarisation of scattered light, since they are almost purely
absorbing. They  serve to reduce the albedo of the silicate mixture to
0.61 at $\lambda=0.22~\mu$m,  where it would otherwise exceed 0.9 for
a mixture consisting only of reflective substances such as silicates or
water ice.

In the denser environment of the circumstellar envelopes of YSOs, the
grains may well be larger than in the dark cloud medium. The parameter
space of size, axis  ratio, composition, deflection angle and grain
orientation was explored with ampld.f code to find grain mixtures
which were capable of reproducing the high CP observed by Bailey et al.(1998)
and Menard et al.(2000) at 2.2~$\mu$m via single dichroic
scattering, while also producing the observed high LP
($>60\%$ at 3~$\mu$m) (Minchin et al., 1991a) and a non-negligible albedo ($\gtsimeq
0.2$) at  2.2~$\mu$m. It was found that for power law index k=3.5, the
grain axis ratio must approach 3:1 and the grain size must extend
up to 0.75~$\mu$m to produce CP of 17\% or more over at least a modest
range of grain orientations and scattering angles. This
upper size limit is in close agreement with the value of 0.8~$\mu$m
independently suggested by Minchin et al.(1991b) by analysing the linear
polarisation in the BN-KL region of Orion. Including even larger
grains reduces the CP produced by the mixture, since in such grains
the phase difference between the parallel and perpendicular
components of the scattered radiation field exceeds 180$^{\circ}$,
causing a CP reversal. The 3:1 axis ratio is larger than is
usually assumed in astrophysical nebulae, but this may be part of the
reason for the very high observed CP. Alternatively, the axis ratio may be
smaller than 3:1 if dichroic extinction rather than dichroic scattering is
responsible for the observed CP (see Section 4). However, the aim of this paper 
is to test the physical feasibility of producing high UV CP rather than to 
calculate the typical values of CP in the radiation field of dark clouds. A
large axis ratio aids the production of high CP by both mechanisms. 
Chrysostomou et al.(2000) found that for perfectly aligned grains a 2:1 axis ratio 
was sufficient to produce the {\it average} value of CP ($\sim 10\%$) in the BNKL-SEBN 
nebula in Orion, which we confirm (and up to 17\% if the combination of grain 
orientation and scattering angle is precisely optimal). A preliminary calculation by 
Bailey et al.(1998) showed that in the case of extremely absorptive grains (n=1.5-0.4i) 
an axis ratio of 2:1 would be also sufficient to reproduce the observed 17\% CP over a 
wide range of grain orientations and scattering angles. However, our calculated 
albedo for that grain mixture is only 0.02 at 2.2~$\mu$m, which is inconsistent 
with the brightness of the observed nebula.

We define two different mixtures of these relatively large grains for use in the 
envelopes of YSOs.

$\bullet$ the D2 mix: $0.005 < a < 0.75~\mu$m, k=3.5, n=1.5-0.2i.

$\bullet$ the SACS2 mix: identical to the SACS1 mix except that the
silicate component has a size distribution $0.005 < a <
0.75~\mu$m. This mix has an albedo of 0.60 at 0.22~$\mu$m.

For completeness we note that different values of k do not appear to increase CP. 

\subsection{Scenario 1 - a single high mass YSO}

	The first scenario that we model is the case of a single
intermediate to high mass YSO, in order to examine the CP
produced in the external radiation field, which would be experienced by a
passing low mass YSO. High mass stars are not thought to be suitable
sites for life (stellar lifetimes $\le 100$~Myr are too short for
terrestrial style evolution and have comparable duration to the
likely period of intense asteroid bombardment) but they dominate the
UV radiation field in star formation regions (see Bailey et al., 1998). This is 
demonstrated by the observation of photoevaporating 'proplyd' envelopes 
around low mass stars, with cometary tails pointing away from the massive O-type
stars at the centre of the Trapezium cluster (eg. O'Dell \& Wen 1994). 
In the case of the BN-KL region there are dozens of low mass stars within a
small fraction of a parsec which could be undergoing asymmetric
photolysis if the radiation field has significant net CP.

The main features of a YSO are illustrated in Figure 4: Embedded YSOs
with ages of order 10$^5$~yr consist of a central protostar surrounded by
a relatively thin accretion disk of gas and dust a few hundred  AU in
extent, which is itself surrounded on scales of up to several thousand
AU by a large envelope which is slowly gravitating toward the disk and
is optically thick even at near infrared wavelengths. The central protostar
contains at least 90\% of the mass of the system (Andre \& Montmerle 1994) 
and the accretion disk has a central hole about the star extending out
to several  stellar radii in which the dust has sublimated (Hamann \& 
Persson 1992; Hillenbrand et al., 1992), with the
consequence that  only a small fraction of the flux from the star is
intercepted by the disk. The envelope contains a bipolar cavity
centred on the star, which is cleared by the stellar wind and allows
light from the star to escape the system along  angles close to the
rotation axis of the system with little or no extinction  by
dust. Hence the intensity and polarisation state of light escaping the
system is a strong function of polar angle. Note that the embedded phase
is much briefer in the most luminous YSOs, for which the above description
may not apply.

A passing YSO which is directly illuminated by the central star along
a line close to the rotation axis will receive a large unpolarised UV
flux, which will efficiently destroy prebiotic molecules and leave no
enantiomeric excess. However, a YSO passing close to the equatorial
plane receives a much lower flux  which is both linearly and
circularly polarised.

\begin{figure*}[thbp]
\begin{center}
\begin{picture}(200,180)

\put(0,0){\includegraphics{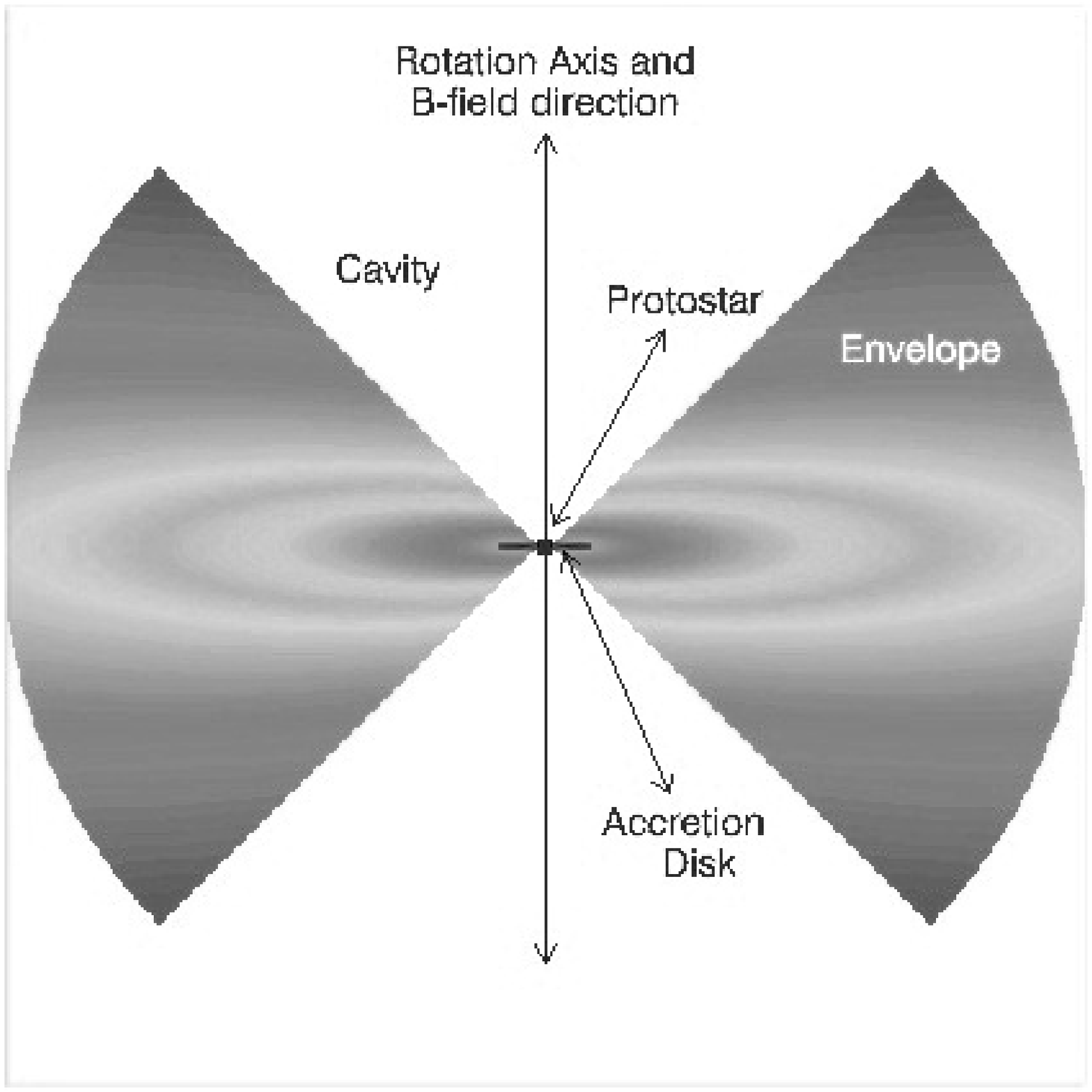}}

\end{picture} 
\end{center}
\small
Figure 4: Schematic diagram of a Young Stellar Object. In this example the
magnetic field is axial and parallel to the rotation axis of the system. The 
density of the envelope declines monotonically with distance from the 
protostar and distance from the disk plane. The density spans several 
orders of magnitude, as indicated by the wrapped greyscale shading. The 
cavity is taken to be evacuated.
\end{figure*}

\subsubsection{Axial Field}

The results of a simulation at $\lambda=0.22~\mu$m with the SACS2 mix are 
illustrated in Figure 5(a-f), for the case of an axial magnetic field. A 
B-field which is axial on scales of thousands of AU is the expected result 
because the 
gravitational collapse of molecular cloud cores with even quite low levels of 
ionisation goes preferentially along the field lines. An axial field has 
previously been observed on scales of thousands of AU (eg. 
Tamura, Hough \& Hayashi 1995) but the observational database suggests that on 
smaller scales the field may be 
axial in some YSOs and toroidal in others (Wright 1994; Smith et al., 2000).

In this example the cavity is conical and has an opening angle of 45$^{\circ}$ to the 
vertical. The accretion disk is a flared structure in vertical hydrostatic equilibrium 
(see Shakura \& Sunyaev 1973) with radius 400~AU, and density given by:\\

(6) $\mathrm{\rho_{disk} = \rho_{0}(R/R_{*})^{-15/8}exp[-0.5(z/h)^{2}]}; 
\hspace{4mm} 400~AU>r>20~R_{*}$

\hspace{1.4cm} $= 0; \hspace{4mm} r<20~R_{*}$

\hspace{1.4cm} $= 0; \hspace{4mm} r>400~AU$\\ 

where $R=\sqrt{r^2+z^2}$ (cylindrical polar coordinates defined by the disk 
plane), $R_{*}$ is the stellar radius, set at 2 solar radii (or $\sim 0.01$~AU), 
and h is the disk scale height, given by:\\

(7) $h = h_{0}(R/R_*)^{9/8}$; $h_0 = R_{*}/75$\\

The size of the inner hole is taken from results in Chiang et al., 2001).
Eqs.(6-7) describes the disk in all YSO simulations in this paper. In fact it has 
little influence on the output. The structural features which do affect model 
outputs most are the surrounding envelope, the system inclination and the 
evacuated cavity. 

In this example the envelope density is given by:\\

(8) $\rho_{env} = C/R^{1.5} [1/((|z|/R)^{v} + 0.05)]; \hspace{4mm} 50<r<5000~AU$

\hspace{1.4cm}        $  = 0; \hspace{4mm} r<50~AU$\\ 

$C$ is a free parameter governing the optical depth from the star to the edge
of the system, on lines of sight which do not intersect the accretion disk.
In the results shown in Figure 5, $C$ is set to a value which leads to optical depth 
$\tau_{60{^\circ}}=3.4$, defined at inclination $i=60^{\circ}$.
$v$ is a parameter decribing the vertical gradient of the envelope, and was set
to 2.0 in the results shown here, so that the envelope is a flattened structure as 
shown in Figure 4. 
Eq.(8) is a simple, empirically derived formalism, adopted 
following the discovery that the $R^{-1.5}$ power law seems to extend throughout 
the envelope in at least some YSOs (Lucas \& Roche 1998). 

\begin{figure*}[thbp]
\begin{center}
\begin{picture}(200,480)

\put(0,0){\includegraphics{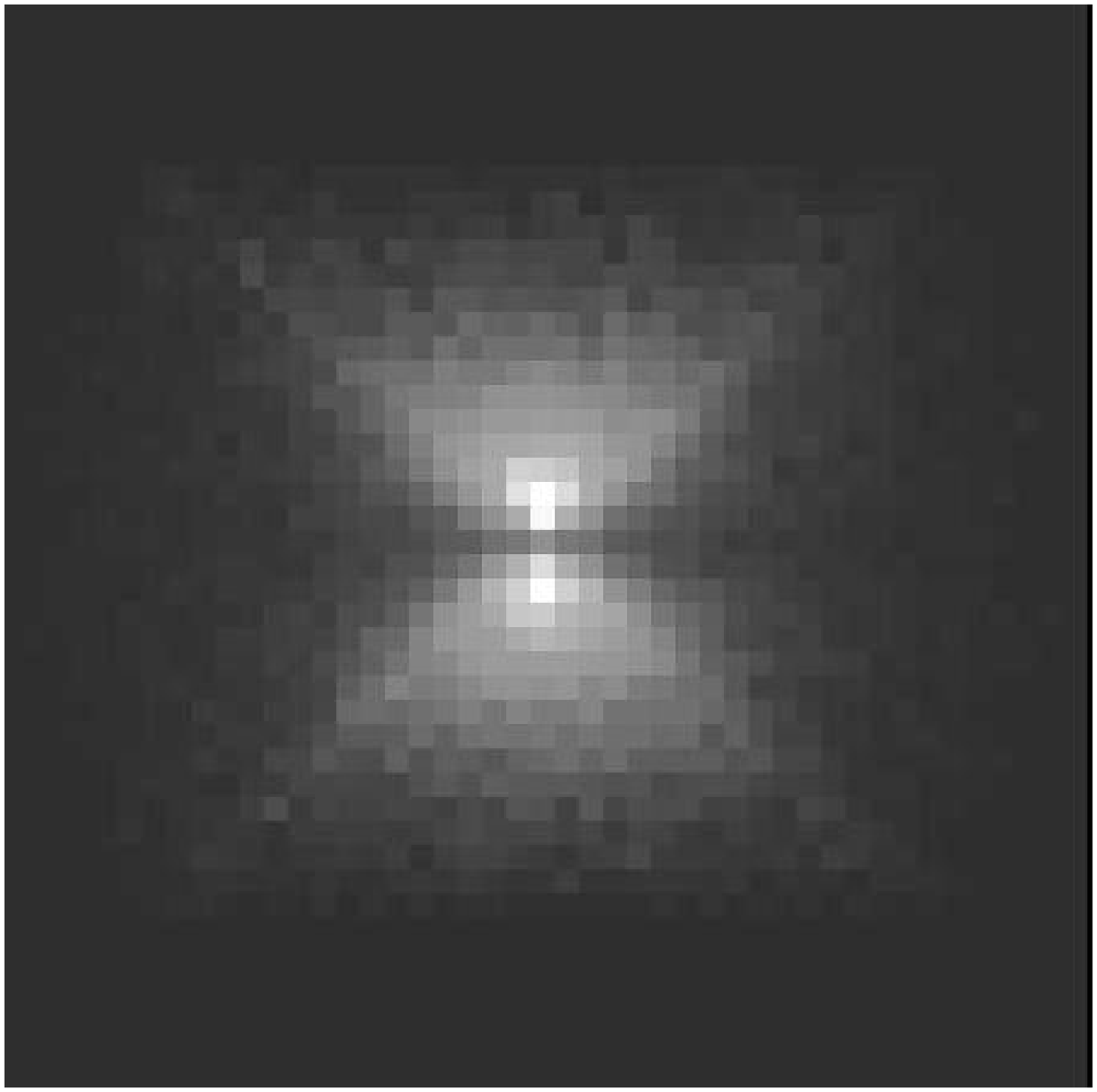}}

\put(0,0){\includegraphics{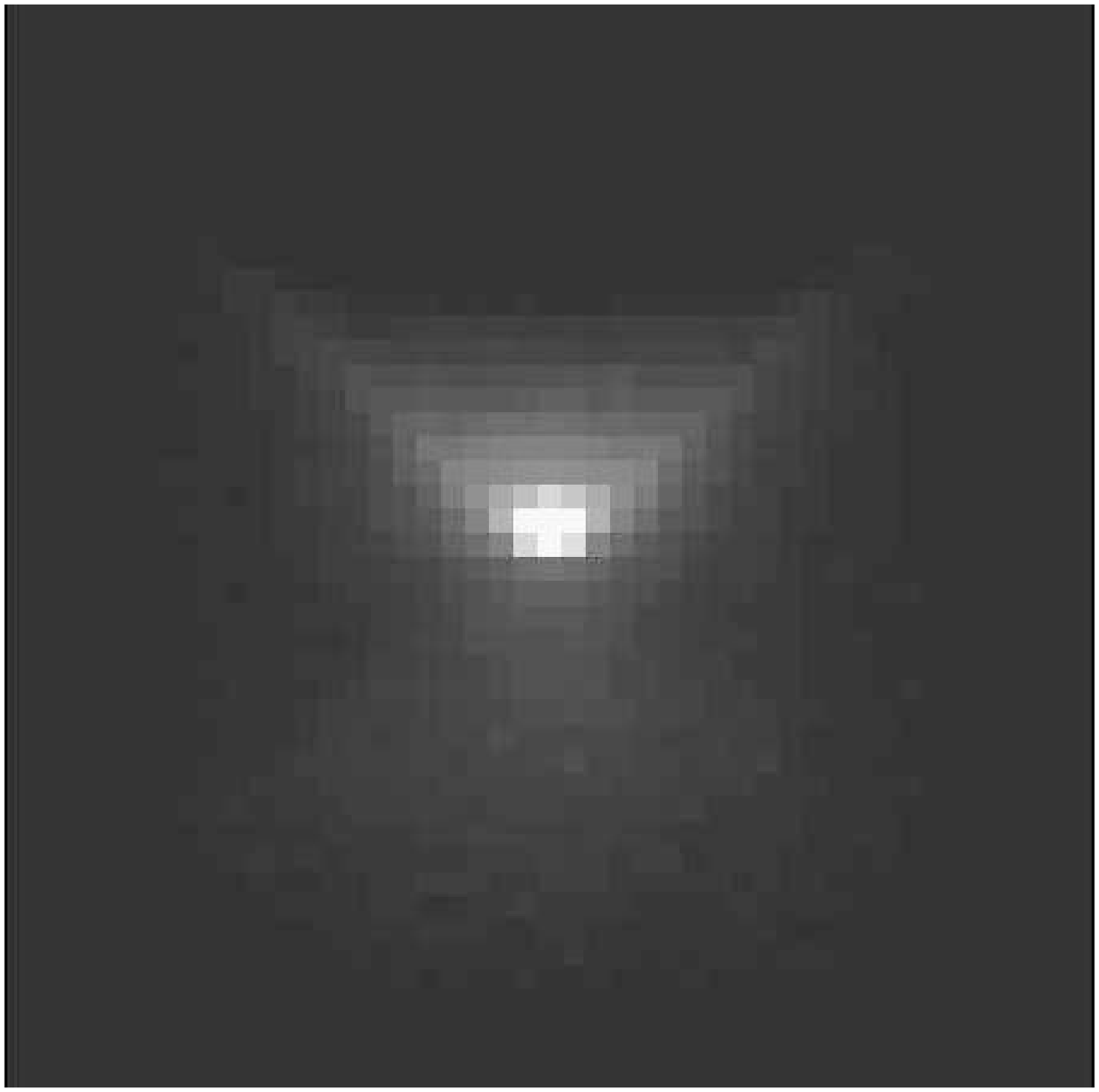}}

\put(0,0){\includegraphics{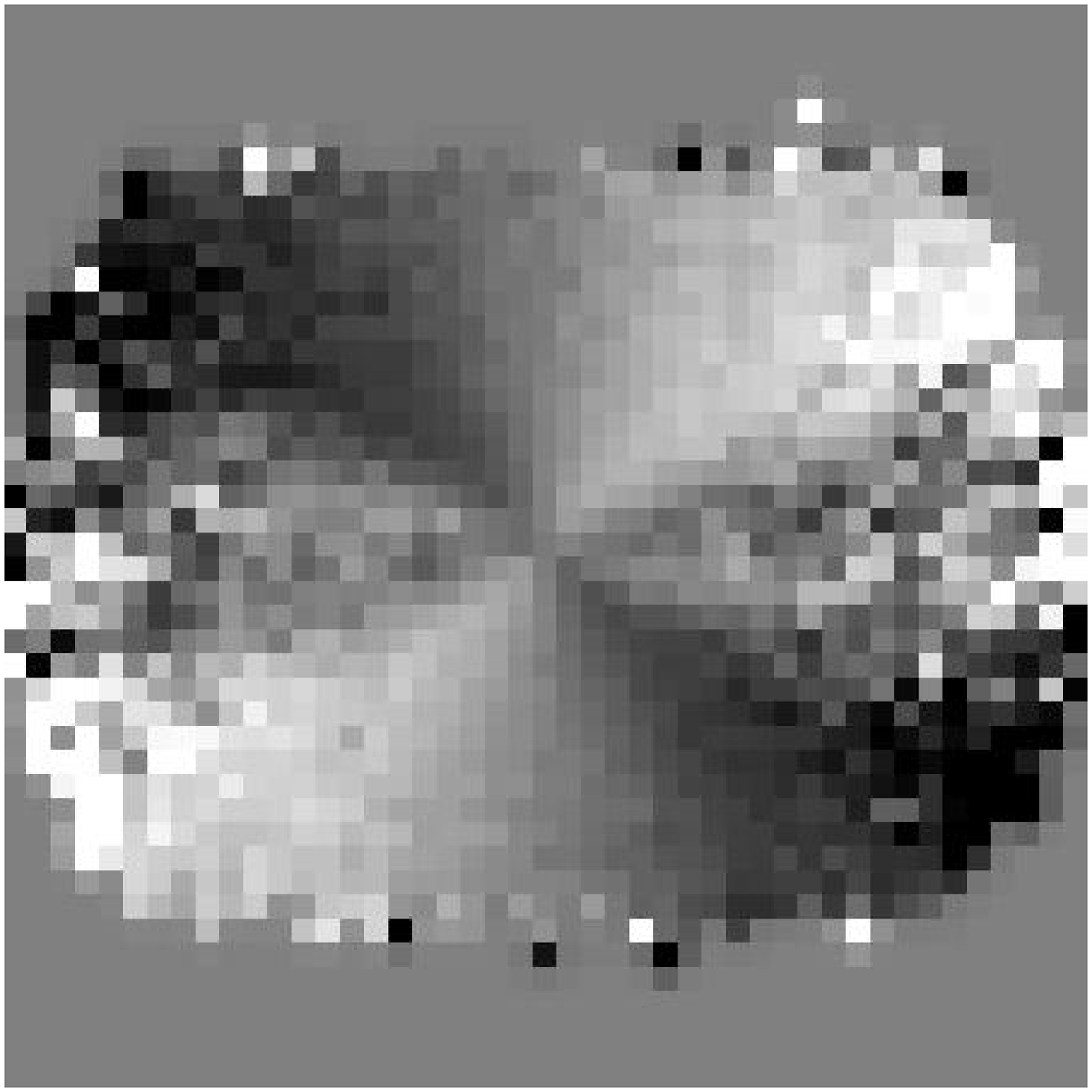}}

\put(0,0){\includegraphics{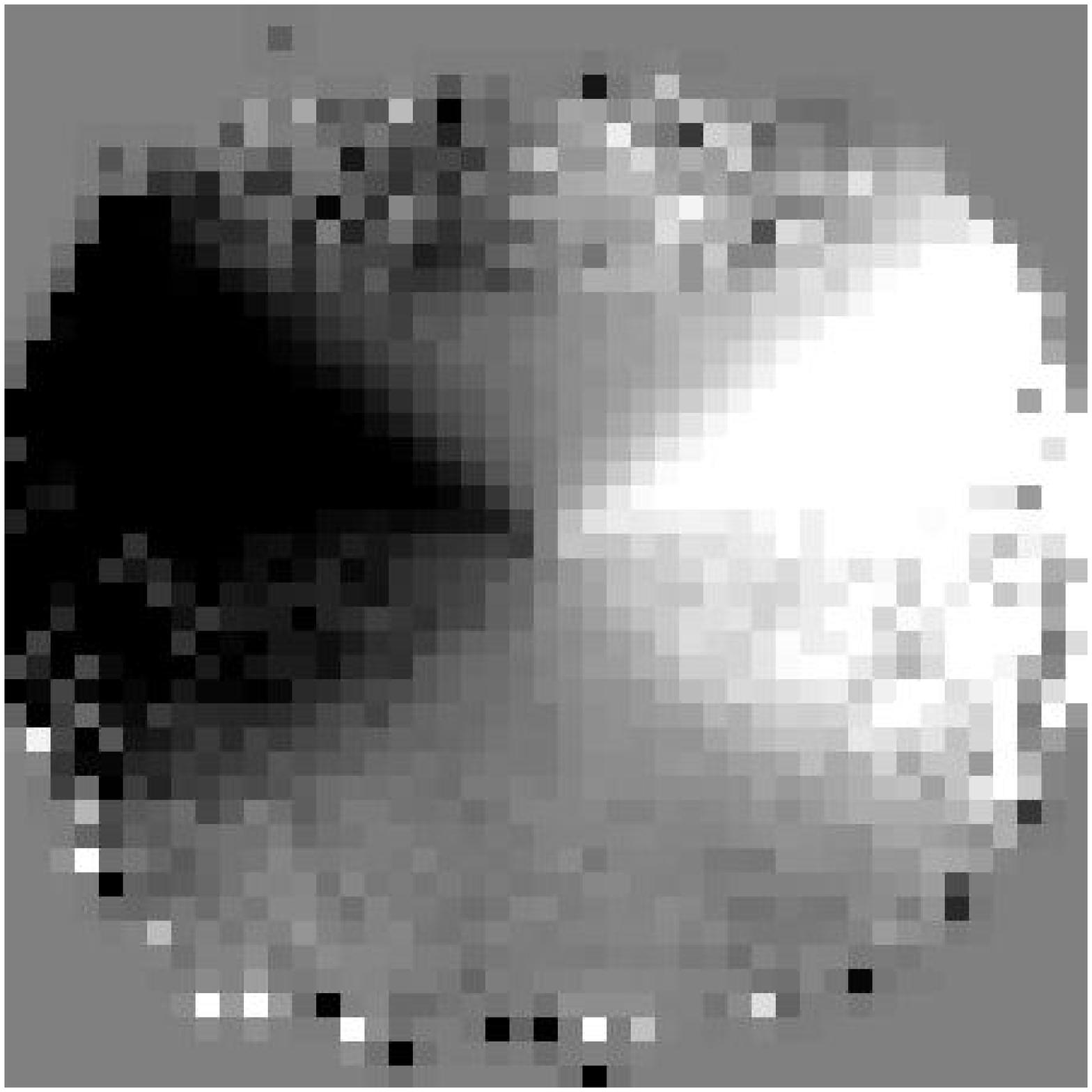}}

\put(0,0){\includegraphics{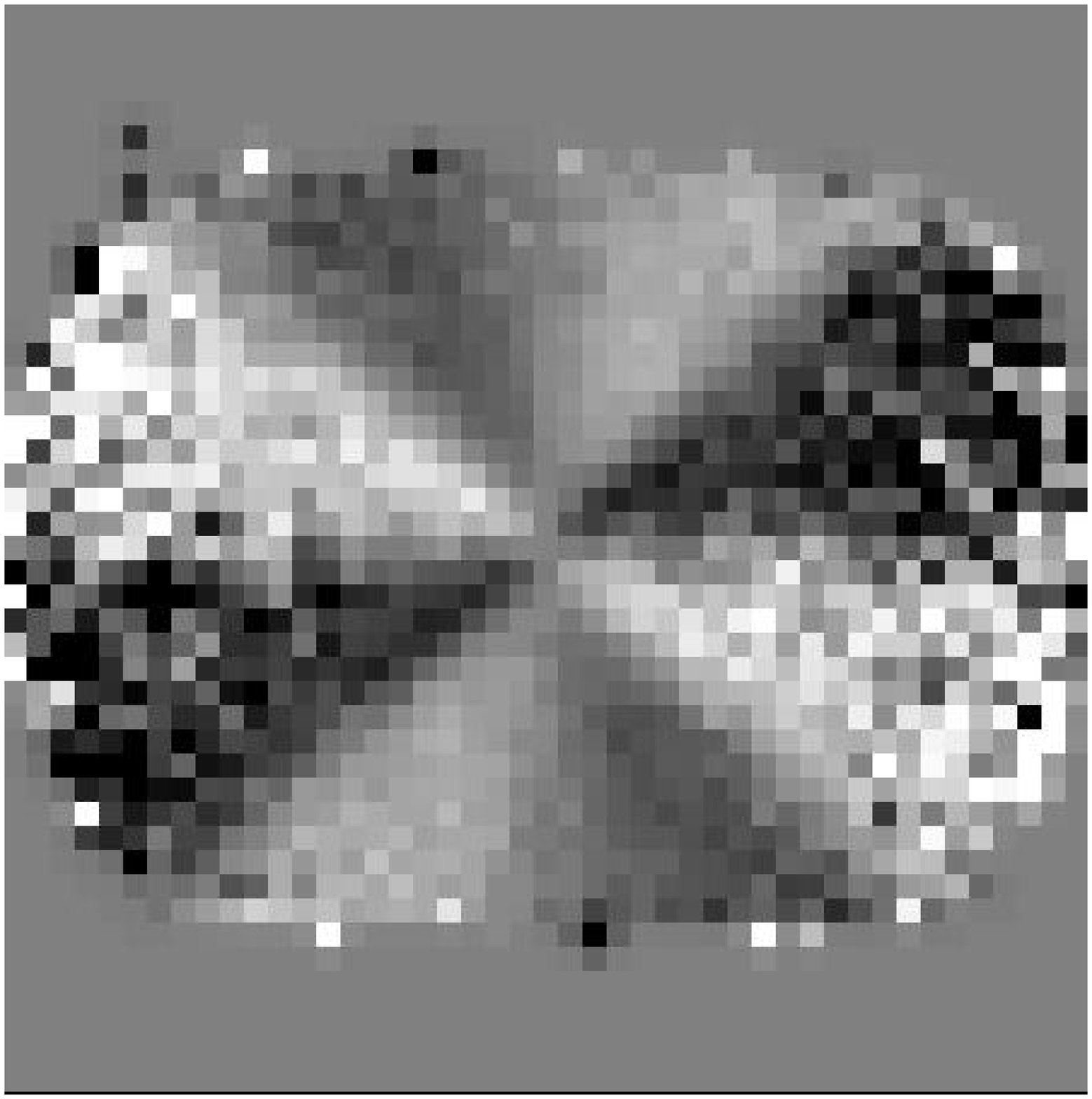}}

\put(0,0){\includegraphics{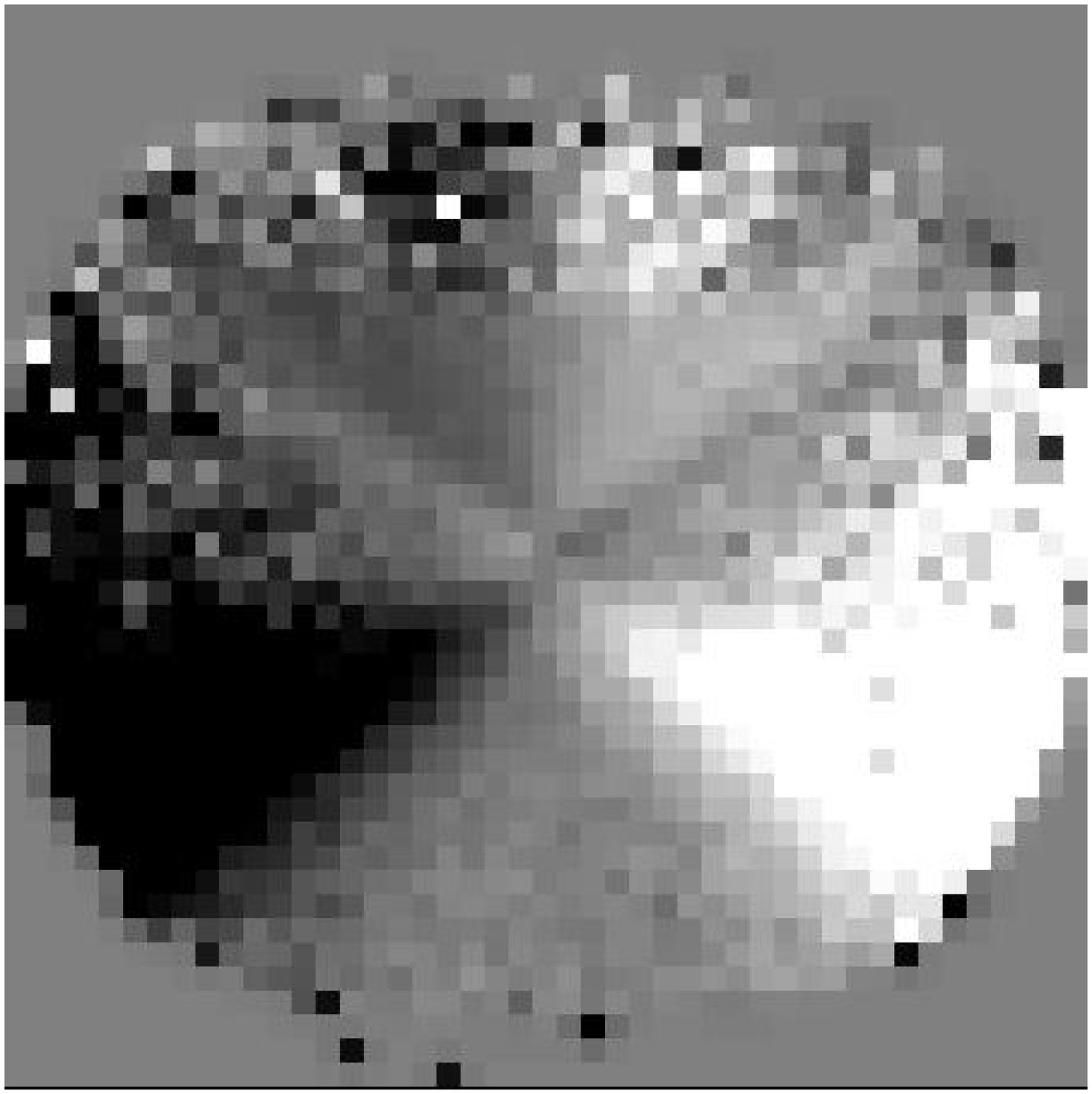}}

\end{picture} 
\end{center}
\vspace{1.4cm}
\small Figure 5. Simulation of a YSO with an axial magnetic field as
viewed in the far field. ($\lambda=0.22~\mu$m, SACS2 dust mix, system
radius=5000 AU). Plots on the left are at $i=90^{\circ}$ and plots on
the right are at $i=72.5^{\circ}$. The top two panels show the flux
distributions (Stokes I) and the rest show CP, scaled from 10\% RCP (dark),
to 10\% LCP (light). The middle panels show the CP produced by dichroic
scattering alone. The bottom panels show the CP after including the effect
of dichroic extinction as well. Regions where the flux is low appear noisy 
in the CP maps.

\end{figure*}

The top panels of Figure 5 show the total flux from the system as seen
in the far field, at inclination angles $i=90^\circ$ and 72.5$^\circ$.
At 90$^\circ$ (viewing in the equatorial plane) the system appears as a
bipolar nebula, with the central star entirely obscured. At $i=72.5^\circ$
($cos(i)=0.3$) the star is still obscured but the image is dominated by light
scattered through small angles in the inner parts of the envelope.
The middle panels of Figure 5 show the CP produced by scattering alone,
and the lower panels show the CP after including the effect of dichroic
extinction. The left-middle panel shows a quadrupolar symmetry: the
sign of CP alternates between adjacent quadrants of the image. This has been 
observed in several YSO nebulae (eg. Chrysostomou et al., 1997) and is simply due to 
alternations in 
the sign of the Stokes U component (part of which is converted to Stokes V) as 
perceived in the scattering plane. The bottom panels show that dichroic 
extinction significantly modifies the CP, even reversing the sense of it
in places where the optical depth is high (close to the equatorial plane).

High CP occurs in parts of the images at both $i=90^\circ$ and 72.5$^\circ$
but not in the regions where the image is brightest. In this example a conical
cavity with an opening angle of 45$^{\circ}$ was used, in order to ensure 
adequate illumination of the regions at fairly low latitude where CP is highest 
in the axial field case (due to the large inclination of the grain short 
axis to the direction of incident photons from the central source).

It is clear the net CP anywhere in the radiation field is zero due to
the reflection symmetry of the system about the rotation axis (z-axis).
We will call this bilateral symmetry, since this term is often used to describe
the symmetry of the human body. Many YSOs have asymmetric nebula, in some 
cases with one side of the nebula being
much brighter than the other (eg. Ageorges et al., 1996; Gledhill et
al. 1996; Stapelfeldt et al., 1999).  This may be due to inclination of
the inner accretion disk (Gledhill 1991), rotating bright spots on the
surface of the star (Wood et al., 2000) or non uniform obscuration by dust in
the cavity or in the foreground. To maximise the net CP, we consider the extreme 
case in a simple manner by summing the model circularly polarised flux from two
diametrically opposite quadrants of the nebula. (The quadrants are
divided by the equatorial plane of the system and the perpendicular
plane containing the axis of the system).  In Figure 6 we plot the net
CP of this idealised asymmetric nebula as a function of wavelength 
received at points in the near field
located 10000 AU from the centre of a system of  radius 5000 AU. The
wavelengths plotted are those corresponding to the standard astronomical 
passbands (UBVRIJHK) and 0.22$~\mu$m. 

\begin{figure*}[h!]
\begin{center}
\begin{picture}(200,250)

\put(0,0){\includegraphics{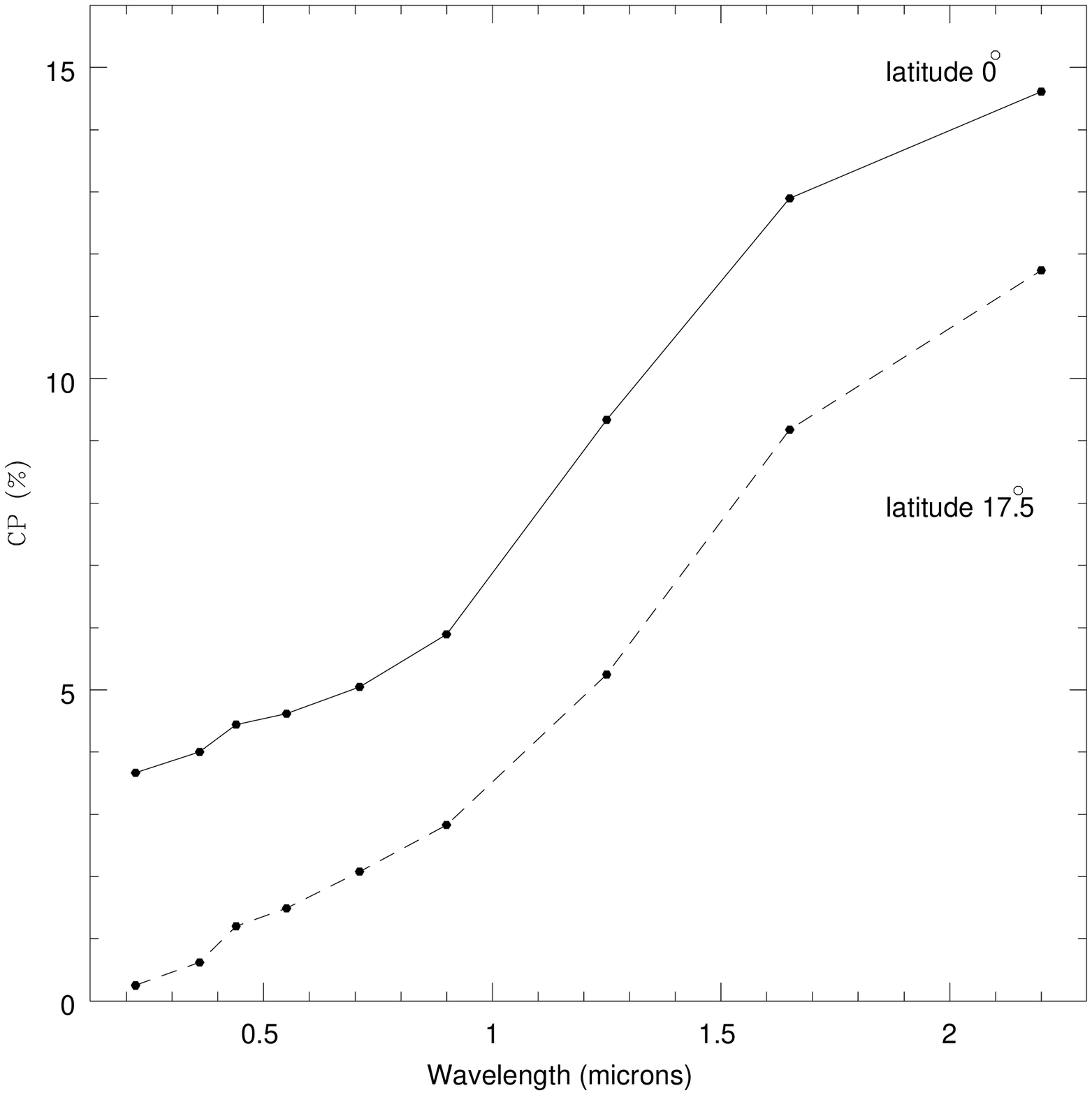}}

\end{picture} 
\end{center}
\small
\vspace{-1.5cm}
Figure 6. Wavelength dependence of circular polarisation in the vicinity
of an YSO with an axial magnetic field. A bilaterally symmetric nebula 
has zero CP at all wavelengths, so here the CP is summed for two diametrically
opposite quadrants of the structure depicted in Figures 4 and 5, providing
a simple representation of the CP from a highly asymmetric nebula (see text).
The calculation is for points located 10000 AU from the centre of a system 
with radius 5000 AU, using the SACS2 mix. The optical depth of the YSO 
envelope is independent of wavelength. Solid line: latitude 0$^{\circ}$ 
(i.e. in the equatorial plane). Dashed line: latitude 17.5$^{\circ}$. At 
latitudes $>17.5^{\circ}$ the CP is greatly diluted at all wavelengths by 
direct flux from the star in this model.

\end{figure*}

    It is important to note that we have chosen to model a bipolar structure with 
constant optical depth (opacity $\times$ density) at all wavelengths, since the 
strong wavelength dependence of grain opacity would otherwise lead to results which 
are uninteresting 
across most of the 0.22~$\mu$m to 2.2~$\mu$m wavelength baseline. Very low optical 
depths lead to an unpolarised radiation field dominated by the central protostar, 
while very high optical depths lead to negligible flux escaping the system.
Since grain opacity increases from the infrared to the UV,
this means the envelope density is lower for the shorter wavelength 
calculations, in order to preserve the product of opacity and density.
We note that only the youngest YSOs, or protostars, display prominent
nebulae in the infrared but many of the less dense envelopes of older
YSOs (eg. classical T Tauri stars) are expected to appear as 
circumstellar nebulae in the ultraviolet. Effectively, we are modelling 
older YSOs in the ultraviolet, and younger YSOs in the infrared.

Exploration of the near field
shows that everywhere the net CP $<5\%$ at 0.22$~\mu$m. This remains
true if the flux from just one quadrant or from 3 quadrants is
included, or if the more absorptive D2 dust grain mix is used instead of
SACS2. Exploration of the envelope parameter space in $C, v$ and cavity opening
angle indicates that YSO envelopes with axial fields cannot produce high CP.

\subsubsection{Helical Field}

	In an attempt to increase the CP in the brighter parts of the
nebula and avoid the problem of zero net CP in bilaterally symmetric systems, 
we model a YSO with a helical magnetic field pitched at 45$^{\circ}$ to the 
equatorial plane. This is merely a test model,
since such a structure is  probably unrealistic, but it is easier to
understand than the more complicated  field structures modelled in the
next section. The helical field breaks the bilateral symmetry of the structure
because there is a choice between a clockwise or anti-clockwise twist. The system
remains axisymmetric (i.e. has rotational symmetry about the z-axis) and is modelled 
with a 2-D code. An anti-clockwise twist is adopted, as viewed from above the 
disk plane.

	The helical field may be expected to produce high CP in parts
of the envelope located at high latitudes via dichroic extinction of
the scattered  light. Scattering at high latitudes causes linear
polarisation mainly in Stokes Q (negative Q by our definition) but in
the frame of reference of a grain inclined by 45$^{\circ}$ to the
rotation axis of the system this is Stokes U. Dichroic extinction of
the scattered light will therefore produce CP.  A narrow cavity
opening angle of 25$^{\circ}$ to the vertical was used, in order to
confine the large amount of flux scattered at the cavity walls to high
latitudes.  For this simulation with the SACS2 mix, dichroic
scattering produces relatively  low CP in the high latitude regions,
since the relevant grain orientations and  deflection angles are not
optimal.

	In Figure 7 we show the results of a simulation of a YSO with
a dense envelope ($\tau_{60^{\circ}}=22$, $v=1.0$) viewed at $i=72.5^{\circ}$. 
These parameters were necessary to create sufficient optical depth in high latitude 
regions to produce CP by dichroic extinction. The other envelope parameters remain
unchanged. The left panel shows the flux
distribution seen in the far field, which shows a bipolar nebula in
which the dense equatorial regions are entirely hidden from view. A
bipolar structure is fairly common among the youngest, most deeply
embedded YSOs, such as NGC6334V (Menard et al., 2000). The brightest
part of the nebula lies along a diagonal line delineating the cavity wall
on the right hand side of the image, due to the larger scattering
cross section for photons travelling in that direction from the
central star.

In the right hand panel is the CP map, which displays the high CP
expected at high  latitudes. The CP is highest in the lower (receding)
lobe. The net CP produced by this model 0.22~$\mu$m in the near field
reaches 12.8\% in the equatorial plane (latitude zero), at 10000 AU 
from the centre of the system. However, the net CP declines at higher  
latitudes, to 6.1\% at latitude 17.5$^{\circ}$ and only 2.2\% at latitude
45$^{\circ}$. For models with a lower envelope density, as in Figure
5, a helical field leads to CP $<5\%$ at all points in the near
field. Although the high density model has some success in producing
high CP, this occurs only in a small portion of the radiation field
and the emerging flux level is low. Hence this scenario is not overly
encouraging for asymmetric photolysis of prebiotic molecules passing
through the radiation field.

\begin{figure*}[h!]
\begin{center}
\begin{picture}(200,235)

\put(0,0){\includegraphics{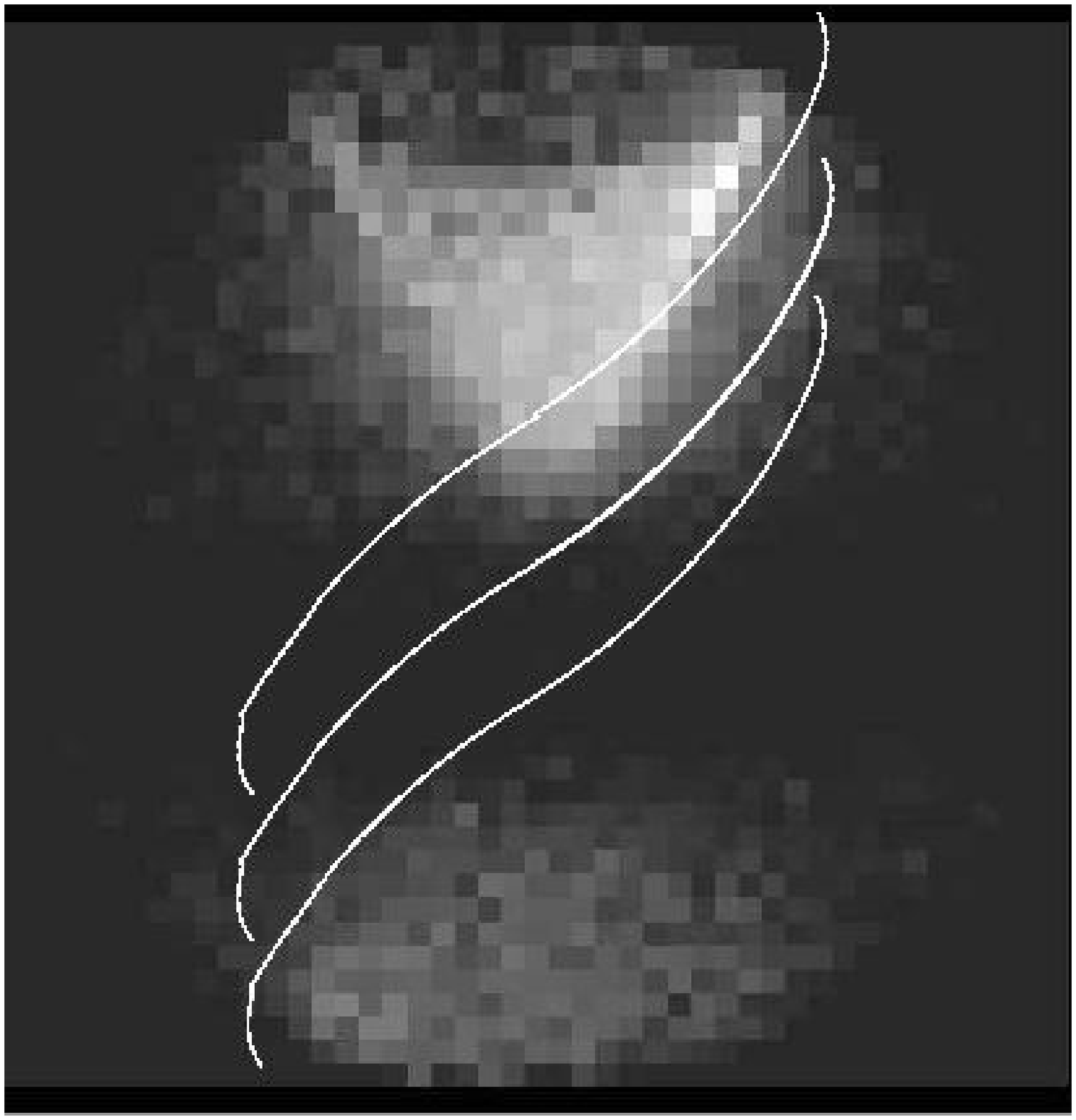}}

\put(0,0){\includegraphics{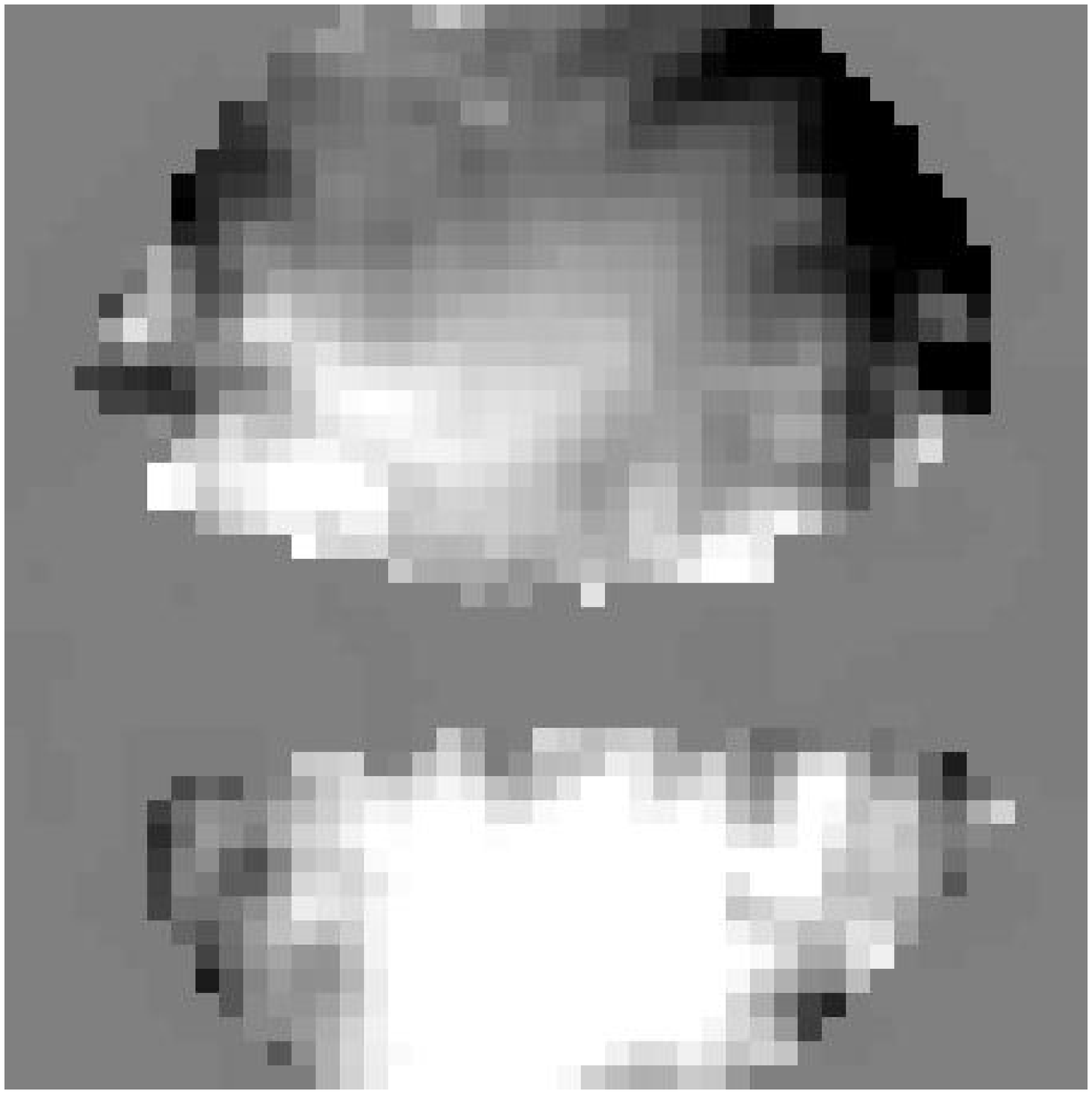}}

\put(0,0){\includegraphics{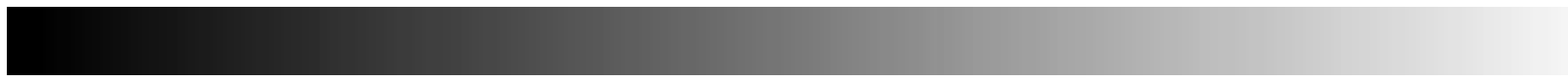}}
\put(80,0){-10\%}
\put(270,0){10\%}
\end{picture} 
\end{center}
\small
Figure 7. Simulation of a YSO with a dense envelope and a helical field,
viewed at $i=72.5^{\circ}$. (left) image, with some illustrative field lines 
superimposed; (right) circular polarisation map. Dichroic extinction 
produces high CP at high latitudes.
\end{figure*}

\subsubsection{Pinched and Twisted Field}

	The axial field and helical field models relied upon scattering
to produce high CP either (a) directly, via the \textbf{Z}$_{41}$ Stokes matrix
element or (b) indirectly, by producing high LP which is
subsequently converted to CP by dichroic extinction (\textbf{K}$_{34}$
extinction matrix element). These models produced little net CP largely
because the grain mixtures are highly forward throwing at 0.22~$\mu$m:
most of the flux is deflected through small angles, leading to little CP
or LP by scattering (see section 2.1).

	In principle, CP can be produced by dichroic extinction alone, if the
magnetic field is twisted (see section 2.2). There are good reasons for 
presupposing that the magnetic field in the inner part of a YSO system is both
pinched and twisted: field lines anchored to the rotating inner accretion
disk will naturally become twisted and inside-out gravitational collapse of 
the parental molecular cloud core will pinch the field, so that it becomes 
stronger near the central star. In this scenario, the optical density of the 
envelope can be reduced since direct flux from the central source will become 
circularly polarised. This scenario is also interesting since the total amount of 
circularly polarised flux may be higher than in the previous scenarios and the 
absence of a diluting unpolarised central source removes the restriction to 
points at low latitude.

The spatial scale and manner in which the putative
pinch and twist occurs is unknown, so we have chosen to parameterise 
the magnetic field in the natural cylindrical polar coordinate system,
in order to explore a variety of pinched and twisted structures.

The adopted formalism is:\\

(9) $B_r =  b_1 z/|z| r^m exp(-(r^n + a|z|))$\\

(10) $B_z =  b_0 + b_1/a r^{m-1} exp(-(r^n + a|z|)) (m+1-nr^n)$\\

(11) $B_\phi = b_2 z/|z| (1/(r^q+az^q+c))$\\

where $b_0$, $b_1$, $b_2$, $a$, $c$, $n$, $m$ and $q$ are positive constants 
which parameterise the field structure. One additional parameter
is the unit of length $L_B$, which was set to values ranging from tens to 
hundred of AU so that the twist occurs on the scale of the inner parts of 
the envelope.

At large radii the field becomes axial ($B_z \rightarrow b_0$, 
$B_r \rightarrow 0$, $B_\phi \rightarrow 0$).
The signs of $B_r$ and $B_\phi$ reverse in the plane of
the accretion disk (z=0) where the pinch and twist is strongest.
A discontinuity in the field direction is permitted at z=0, implying
the existence of a current sheet in the plane of a thin accretion 
disk. The form of $B_z$ relates to $B_r$ in a manner constrained by the 
divergence equation (Div {\bf B}=0). By contrast $B_{\phi}$, the twist
component, is independent of the other components since the twist 
is axisymmetric and therefore does not contribute to the divergence equation. 

An example of the field structure produced by Eqs.(9-11) is shown in 
Figure 8. The parameter space was explored with the Monte Carlo code 
for cases where the field lines in the circumstellar envelope cross 
the disk plane only once. Parameters which might potentially produce
high net CP were identified by calculating the CP of unscattered (i.e.
direct) light from the central star for all system inclinations, $i$.
It was found that structures in which the twist in the field extends 
throughout the circumstellar envelope produce the highest CP in the direct 
beam. These correspond to a small value of q (see Eq.11).
The optical depth of the envelope and the twist of the field are both 
greatest at $i=90^{\circ}$, so the birefringence of the envelope is
greatest in this direction. The maximum of CP in the direct beam
usually occurs at a lower angle, where the integrated phase difference 
between orthogonal electric field components (which gain an initial linear 
polarisation near the cavity walls) reaches $90^{\circ}$.

\begin{figure*}[thbp]
\begin{center}
\begin{picture}(200,280)

\put(0,0){\includegraphics{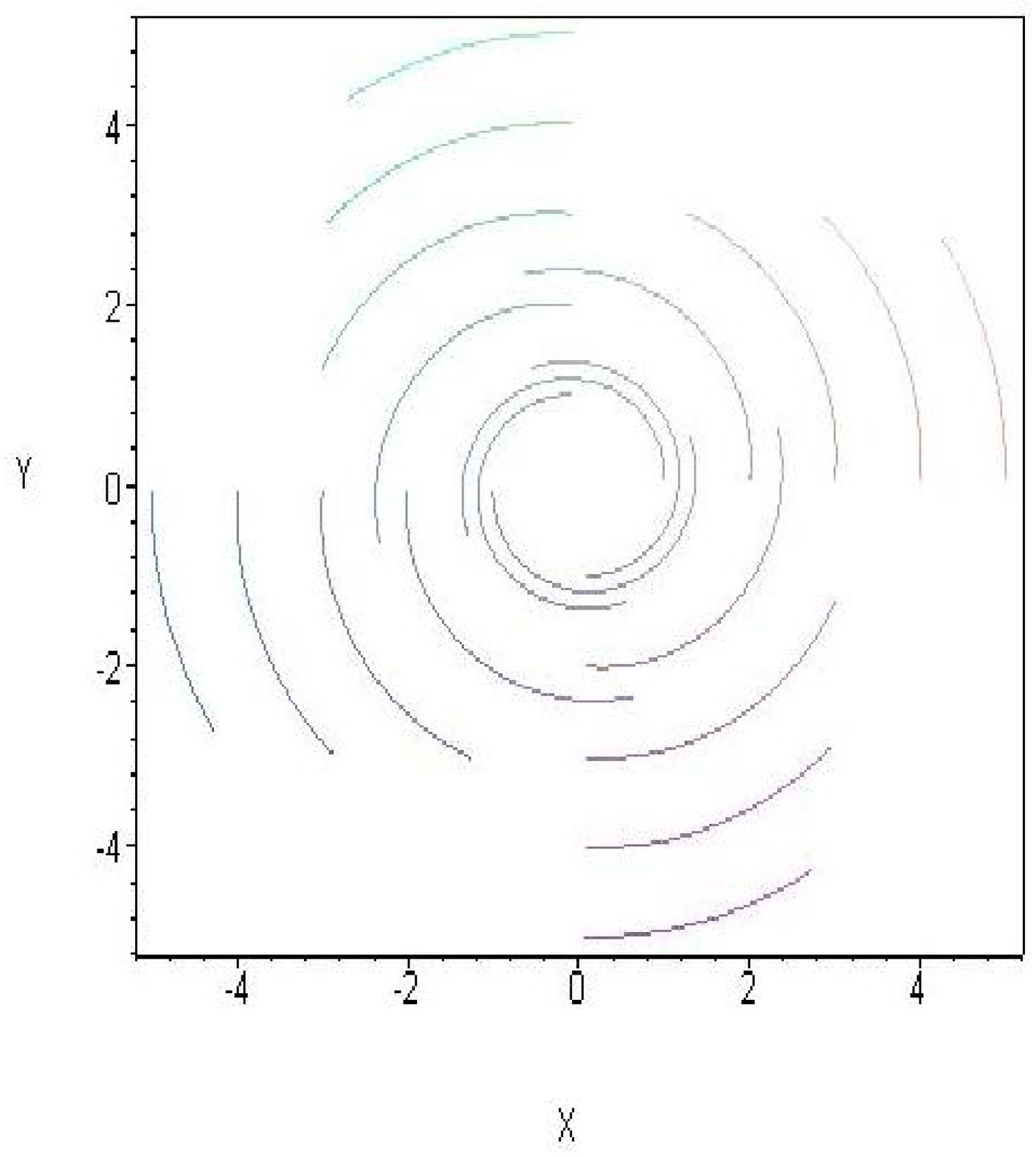}}

\put(0,0){\includegraphics{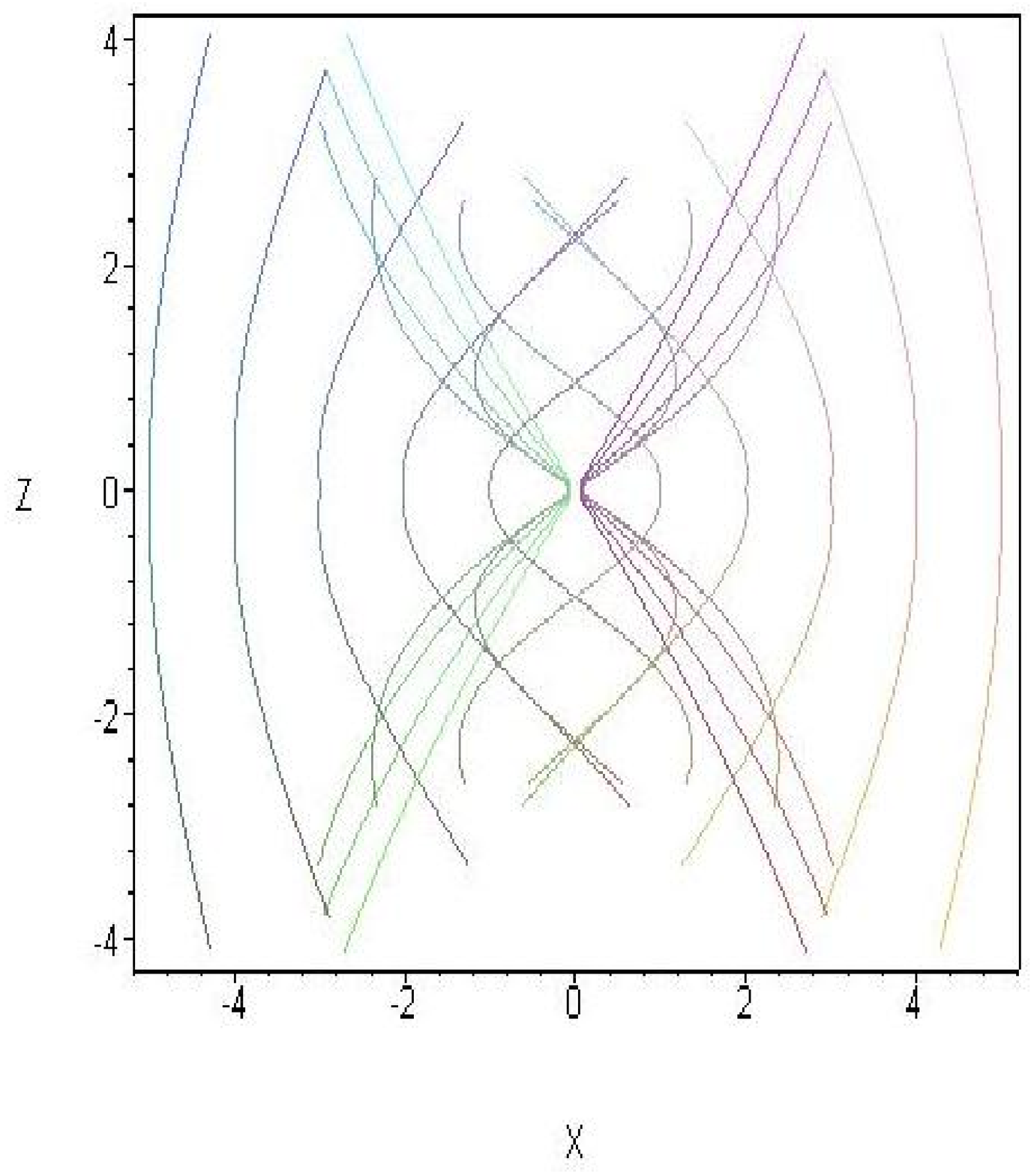}}

\end{picture} 
\end{center}
\small
\vspace{-2cm}
Figure 8. Field structure for (n=1,m=1,q=1,a=0.4,b0=0.5,b1=0.5,b2=2.0,c=0.1),
see Eqs.(9-11). The field lines are highly twisted but only weakly pinched.
(left) view of the field lines looking down on the disk plane. (right) view
from the equatorial plane, showing the pinch more clearly.
\end{figure*}

In Table 1 we list the net CP produced by an axisymmetric system of 5000~AU 
radius with the field structure in Figure 8, as a function of 
inclination. For this model the density structure of Ulrich (1976) is adopted
for the YSO envelope (better known as the Terebey, Shu \& Cassen (1984) inner 
solution for the collapse of a rotating isothermal sphere). This structure is 
similar to Equ.(8) except that the radial density dependence becomes shallower 
than $R^{-1.5}$ at radii $R \ltsimeq 3R_c$ ($R_c$ is called the centrifugal radius) 
approaching $R^{-0.5}$ in the innermost parts of the envelope. This alternative
envelope density function was used since the shallower radial density profile
spreads the region of highest optical density over a larger volume. Hence, the
field is significantly twisted in the dense region where most of the dichroic 
extinction occurs, which aids the production of CP. In this model  $R_c= 100$~AU 
and $L_B = 40$~AU were adopted. 
A conical cavity was used, opening at 25$^{\circ}$ from the vertical, with a 
radius of 80 AU in the disk plane. The calculation is at 0.22$~\mu$m 
for the SACS2 mix.

Net CP is listed both in the near field at 10000~AU from the 
central star, and in the far field. Since the optical depth is relatively low 
($\tau_{60^{\circ}}=3.0$), the CP of the direct beam is 
also tabulated, since this makes a large contribution to the circularly 
polarised flux at most inclinations.
It is apparent that this field configuration produces high CP in the direct 
beam at some inclinations but this is diluted by the contribution from 
scattered 
light with low CP so that total net CP$<5\%$ at all points in the external 
radiation field. If the optical depth is small enough to make dilution by 
scattered light unimportant ($\tau \ltsimeq 3$), then the CP of the direct 
beam is always $<10\%$ for the SACS2 mix. Optical depth $\tau > 2$ is 
required to give the beam strong LP ($K_{12}/K_{11}= 0.266$ for the SACS2 
mix) and a further optical depth of at least unity is needed to convert 
a significant fraction of this to CP, even for an ideal field structure, 
since $K_{34}/K_{11}= 0.156$ for the SACS2 mix.

Extensive exploration of the parameter space (field structure; envelope mass,
size, and density structure; cavity size) has not produced a model in which the
scattered light has a high net CP. This appears to be due to the large number 
of paths
by which scattered photons can reach a given point in the radiation field,
a problem which is in part due to the high albedo (0.6) of the SACS2 mix.
As noted earlier, more absorptive grain mixtures such as D1 and D2 have very 
low values for the $K_{34}/K_{11}$ ratio. Hence, it appears unlikely that 
YSOs with pinched and twisted magnetic fields can produce high net CP at
0.22~$\mu$m via dichroic extinction by sub-micron sized grains.

\pagebreak
\normalsize
\begin{center}
\textbf{Table 1 - Circular Polarisation from a YSO with a pinched and
twisted field}\\
\begin{tabular}{lrrrr}
 & & & & \\
cos(i) & $\tau^a$ & CP (direct beam) & Total CP$^b$ (10000 AU) & Total CP$^b$ (far field) \\ \hline
0.0 & 21200 & -10.1\% & -0.1\% & 0.0\% \\
0.1 & 31.9 & 23.2 \% & 1.0\% &  0.9\% \\
0.2 & 4.6 & 18.1\% & 3.2\%   &  3.1\% \\
0.3 & 3.9 & 12.5\% & 4.2\%   &  4.0\% \\
0.4 & 3.4 & 8.2\% & 4.3\%    &  4.0\% \\
0.5 & 3.0 & 5.0\% & 3.7\%    &  3.5\%\\
0.6 & 2.7 & 2.7\% & 2.9\%    &  2.7\% \\
0.7 & 2.3 & 1.1\% & 2.0\%    &  1.9\% \\
0.8 & 1.7 & 0.2\% & 1.2\%    &	1.2\% \\
0.9 & 0.0 & 0.0\% & 0.7\%    &	0.6\% \\
0.9675 & 0.0 & 0.0\% & 1.0\% &	1.0\% \\
0.9925 & 0.0 & 0.0\% & 1.0\% &  1.0\% \\
\end{tabular}
\end{center}
\small
Notes: (a) $\tau$ is the optical depth to the central star for the given 
inclination.\\
\hspace{6mm}(b) Total CP includes the contributions from direct and scattered light.\\

\normalsize
\subsection{Scenario 2a - two uniform sheets of nebulosity with inclined fields}

	In Section 2.2 we showed how a uniform slab of cloud can produce
CP via dichroic extinction of a linearly polarised beam. The linear
polarisation of the incident beam could arise from scattering or prior 
dichroic extinction. Here we show the results of a Monte Carlo simulation
of the latter scenario: light passing through 2 parallel uniform 
sheets containing dust grains which are perfectly aligned with uniform 
magnetic fields (see Figure 9). The fields in each slab are parallel to the 
plane of the slab but inclined to each other by some angle $\zeta$. The CP 
of the emerging radiation, including the contribution
from scattered light, is listed in Table 2 as a function of cloud optical
depth ($\tau$), distance beyond the second cloud (d) and $\zeta$, using the 
SACS1 grain mixture. This scenario is 
similar to earlier calculations by Martin (1974) but modern computing power 
allows us to include the important effects of scattered light.


\begin{figure*}[thbp]
\begin{center}
\begin{picture}(200,210)

\put(0,0){\includegraphics{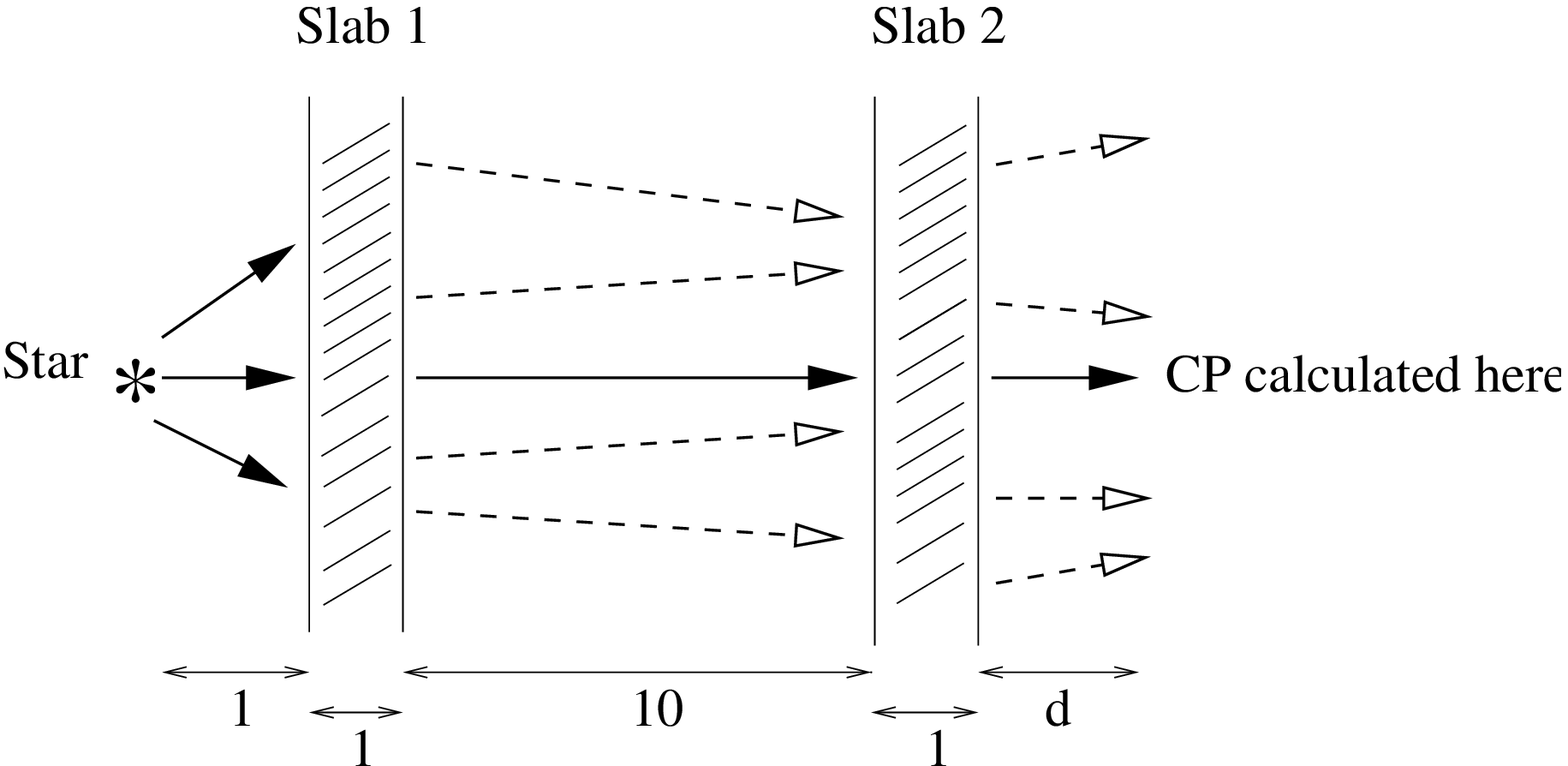}}

\end{picture} 
\end{center}
\small
\vspace{-1.5cm}
Figure 9. Cartoon illustrating Scenario 2a. The direct beam from the star
perpendicular to the slabs is shown as a solid line; the dashed lines 
illustrate rays of scattered light. Both slabs of nebulosity contain 
perfectly aligned oblate spheroids with magnetic fields orthogonal to the 
direct beam. The field directions are inclined to each other by an angle 
$\zeta$ in the plane of the slabs. CP is calculated along the line of the 
direct beam at various distances, d, beyond slab 2. Distance units are 
arbitrary.
   
\end{figure*}

	It is clear that this scenario produces fairly high CP 
($\approx 10\%$) with the SACS1 mix. The total CP is lower than for the 
idealised case shown in Figure 3, since the the first slab produces $<100\%$
LP and scattered light dilutes the CP produced by the direct beam. Scattering 
tends to remove the memory of the incident polarisation state, so multiply 
scattered photons (sometimes scattered several times within a slab) usually have 
lower CP than singly scattered photons or the direct beam in these 
simulations. For the SACS1 mix the typical albedo is 0.61 (albedo varies slightly
with grain orientation), so the average number
of scatterings in an optically thick nebula is $1/(1-0.61)\approx 2.5$.
As noted in Section 3.2.3, the D1 mix produces much lower CP despite its low albedo, 
since the birefringence of more absorptive grains is much lower 
($K_{34}/K_{11}=0.043$, see also Figure 2). The requirements for this scenario to 
produce high CP are therefore: 

\begin{itemize}
\item A well ordered field structure in both of the cloud sheets, the two
clouds having field directions which are significantly different. In the
case of random field orientations, half of all pairs of cloud sheets
would have field inclinations within $22.5^{\circ}$ of $\zeta=45^{\circ}$ or
$\zeta=135^{\circ}$, leading to high Cpol.

\item A grain mixture with a small imaginary component of refractive index
(i.e. a shiny mixture), which is quite likely for a mixture dominated by eg. 
silicates and water ice in star forming regions.

\end{itemize}

\normalsize
\begin{center}
\textbf{Table 2 - Circular Polarisation from Scenario 2a}
\begin{tabular}{lrrrrr}
 & & & & & \\
$\tau$$^{a}$ & $\zeta$ & CP (direct beam) & Total CP (d=1) & Total CP (d=10) & Total CP (d=20)\\ \hline
1	&  0$^{\circ}$  & 0\%	  & 0.0\% &  0.0\% & 0.0\% \\
2	&  0$^{\circ}$  & 0\%	  & 0.0\% &  0.0\% & 0.0\% \\
3	&  0$^{\circ}$  & 0\%	  & 0.0\% &  0.0\% & 0.0\% \\
4	&  0$^{\circ}$  & 0\%	  & 0.0\% &  0.0\% & 0.0\% \\ \hline

1	& 15$^{\circ}$  & 2.52\%  & 0.68\% & 1.33\% & 1.73\% \\
2	& 15$^{\circ}$  & 6.77\%  & 2.10\% & 3.70\% & 4.32\% \\
3	& 15$^{\circ}$  & 9.24\%  & 2.59\% & 5.01\% & 5.55\% \\
4	& 15$^{\circ}$  & 9.64\%  & 3.52\% & 3.93\% & 4.68\% \\ \hline

1	& 30$^{\circ}$  & 4.50\%  & 1.06\% & 2.51\% & 3.06\% \\
2	& 30$^{\circ}$  & 12.9\%  & 3.86\% & 6.56\% & 8.23\% \\
3	& 30$^{\circ}$  & 18.5\%  & 5.13\% & 7.67\%  & 9.52\%  \\
4	& 30$^{\circ}$  & 19.9\%  & 3.69\% & 7.36\%  & 8.39\% \\ \hline

1	& 45$^{\circ}$  & 5.43\%  & 1.64\% & 3.23\%  & 3.59\% \\
2	& 45$^{\circ}$  & 17.1\%  & 5.72\% & 8.65\%  & 9.41\% \\
3	& 45$^{\circ}$  & 27.0\%  & 6.75\% & 10.5\%  & 11.8\% \\
4	& 45$^{\circ}$  & 31.2\%  & 5.39\% & 9.27\%  & 9.33\% \\
\end{tabular}
\end{center}
\small
(a) $\tau$ is the optical depth through each slab for unpolarised light.\\

\normalsize

\subsection{Scenario 2b - a YSO and a cloud sheet}

	The alternative scenario whereby the LP is
produced by scattering is also possible. The D2 mix produces a
maximum LP up to 90\% at 0.22~$\mu$m by single
scattering for deflection angles near 90$^{\circ}$ (with the oblate grains
oriented face on to the incident ray). Similarly, the SACS1, SACS2, and D1
mixes can produce LP up to 76\%, 74\% and 88\%
respectively. However, the difficulty for scattering models remains
the tendency for forward scattering in the ultraviolet, which causes
most of the scattered flux to have much lower LP. To
produce a radiation field with high LP it is
therefore necessary to construct a scenario in  which light from an
unpolarised source (eg. a star) is scattered through $\sim 90^{\circ}$ but
direct flux from the source is somehow obscured, in order to prevent
dilution.

	Scattering in the lobes of a bipolar nebula is one example of
such a scenario. YSO envelopes viewed at $i \approx 90^{\circ}$ appear
as bipolar nebulae with the central star obscured, as shown in Figure
7 for the case of a helical field. The scattered light from a bipolar
nebula has significant LP which is mostly parallel to
the disk plane, provided dichroic  effects are
unimportant. (Competition between dichroic extinction and  dichroic
scattering tends to reduce the degree of polarisation).

To quantify the LP at 0.22~$\mu$m, Monte Carlo simulations of an
axisymmetric system were run for YSOs with spherical grains (so that
there are no dichroic effects), for both low density and high density
envelopes described by Equ.(8). Results are given in Table 3. Grains with 
the size distribution and optical properties of the D2 mix were used to help
produce high LP. The relatively low albedo of the D2 mix, 0.43, means
that the emergent radiation is dominated by singly scattered light,
which tends to have higher LP than multiply scatterd light. A narrow
cavity opening angle of 20$^{\circ}$ to the vertical was used in order
to ensure that most of the emerging flux is polarised parallel to the
disk plane (negative Q), sampling only a small sector of the
centrosymmetric pattern. The resulting flux distributions for the low density
and high density simulations were similar to those shown in the top
panels of Figure 5 and the left panel of Figure 7 respectively.

The results in Table 3 show that LP of up to 30\% is
produced in systems with high optical depth, viewed at $i=90^{\circ}$.
In systems with relatively low optical depth  more flux is able to emerge after
scattering at low latitudes, so the the forward throwing phase
function significantly reduces the LP.
If the YSO has an active molecular outflow, there may be a modest optical depth
of dust in the cavity, which would increase the scattering at high latitudes
with polarisation parallel to the disk plane. After adding a uniform dust component 
in the cavity, with optical depth of 0.75 along the axis from the protostar to the 
edge of the system, the maximum LP was raised from 30.8\% to 39.4\% in the 
$\tau_{60^{\circ}}=36$;$v=1.0$ case.

It is clear that a bipolar nebula producing almost 40\% LP at
$\lambda=0.22~\mu$m could lead to fairly high CP after dichroic extinction 
by an interstellar cloud sheet, as in Scenario 2a. However, there are 
some reasons why Scenario 2(a) might be preferred bover 2(b): first, in 2(b) the highest 
LP is produced in the weakest part of the radiation field around a very
optically thick nebula; second, the LP produced by
scattering in an optically thick nebula is lower than that produced by 
dichroic extinction (see Figure 3).

\pagebreak
\normalsize
\begin{center}
\textbf{Table 3 - far field LP from a bipolar YSO with spherical grains}
\begin{tabular}{lcc} 
 & & \\
Model		&   cos(i)  & LP \\ \hline     
$\tau_{60^{\circ}}=36$;$v=1.0$ & 0.0  & 30.8\%  \\
$\tau_{60^{\circ}}=36$;$v=1.0$ & 0.5  & 18.2\%  \\
$\tau_{60^{\circ}}=5.9$;$v=1.0$ &  0.0  & 17.8\%\\
$\tau_{60^{\circ}}=5.9$;$v=1.0$ &  0.5  & 5.1\%$^a$ \\
\end{tabular}
\end{center}
\small
Note: (a) LP is strongly diluted by flux from the central source at 
$i \ltsimeq 60^{\circ}$.

\normalsize
\section{CP in Orion and implications for grain axis ratio}

In this section we model the high CP measured at infrared wavelengths in Orion, to 
see what insights can be gained for application to UV modelling. 
The BNKL-SEBN region of Orion displays the centrosymmetric pattern of LP vectors normally 
associated with scattering, rather than the aligned vectors associated with dichroic 
extinction. The situation in NGC6334V (the other known case of high infrared CP) is 
unclear, since the nebulous region is too small for a clear LP pattern to be determined. 
There is no doubt that BNKL-SEBN is a reflection nebula but it is quite possible that the 
CP of the radiation from this nebula is due to dichroic extinction of the scattered 
light: at infrared wavelengths linear dichroism has less effect on the polarisation 
state than birefringence (see Figure 2 and Martin 1974). Hence, high CP can be produced 
without much disturbance of the centrosymmetric vector pattern. 

We illustrate this with an infrared model of the BNKL-SEBN region in Figure 10, which 
shows the CP and LP produced by a variation on the SACS1 grain mix, using grains with 
axis ratio of only 1.5:1. For this model a forced scattering code 'shadow.f' was used,
which is described in Lucas et al.(2003). The magnetic field direction in the plane
of the sky has been measured by Chrysostomou et al.(1994) via high spatial resolution
observation of the 2.12~$\mu$m H$_2$ emission in the region. The data indicates that 
the grain aligment axis is tilted by approximately 60$^{\circ}$ to the Poynting vector 
of rays from the illuminating source IRc2. This orientation is suitable for production
of CP by either dichroic extinction or dichroic scattering. In this simple model we 
assume that the magnetic field lies exactly in the plane of the sky and that IRc2 and 
BNKL-SEBN lie at the same distance from the Earth, so that the scattering angle is 
90$^{\circ}$. The light from SEBN is then passed through a uniform slab of cloud
in the foreground with the same field orientation, which amplifies the dichroic 
extinction occurring within SEBN itself. The existence of this foreground extinction is 
inferred from 
the 1.25~$\mu$m data of Chrysostomou et al.(2000), in which the entire region is 
obscured from view (see also Brand et al., 1988). The left panel of Figure 10 shows that 
the observed CP of 15-20\% at 2.2~$\mu$m (see Chrysostomou et al., 2000) is reproduced 
by this model, while the centrosymmetric LP pattern observed by Minchin et al.(1991) 
is also reproduced (right hand panel). The centrosymmetric pattern also acts as a 
constraint on dichroic scattering models, since scattering by highly flattened
grains also produces some alignment of the vectors, perpendicular to the alignment caused 
by dichroic extinction (eg. Whitney \& Wolff 2002; Lucas 2003).

\begin{figure*}[thbp]
\begin{center}
\begin{picture}(200,220)

\put(0,0){\includegraphics{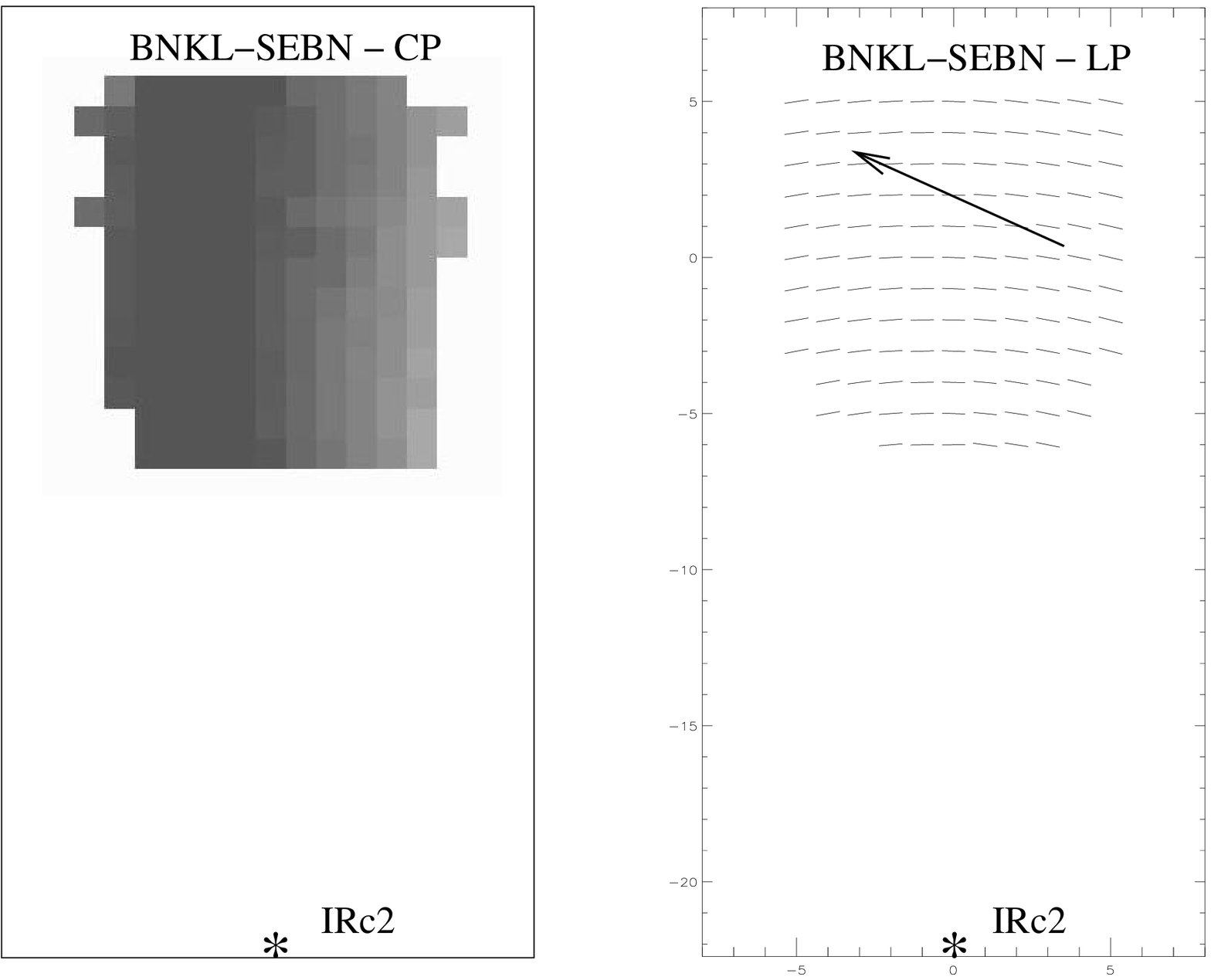}}

\end{picture} 
\end{center}
\small
\vspace{-1.7cm}
Figure 10. A simple infrared model of BNKL-SEBN. In this simple model the nebulous 
cloud BNKL-SEBN is illuminated by the YSO IRc2. Scattering by oblate spheroids with 
axis ratio 1.5:1 reproduces the centrosymmetric pattern (right panel) observed by 
Minchin et al., (1991). Dichroic extinction by a foreground sheet of cloud and BNKL-SEBN 
itself reproduces the maximum observed CP of 15-20\% (left panel) without perturbing the 
LP vectors by more than 2$^{\circ}$. The magnetic field field direction in SEBN and the 
foreground sheet (tilted at 60$^{\circ}$ to the incident rays from IRc2) is indicated 
by the arrow in the right hand panel. 
\end{figure*}

In fact this is an imperfect model since the predicted degree of LP at 2.2~$\mu$m (90\%)
is far higher than the observed value (38\% at 2.2~$\mu$m and 57\% at 3.6~$\mu$m), owing
to the 90$^{\circ}$ scattering angle.
To improve the model we varied 5 parameters: scattering angle, maximum grain size,
grain axis ratio, optical depth and the magnetic field orientation. Three classes of 
successful solution were found, all of which reproduce to a close approximation the 4 
observables described above (CP percentage; LP centrosymmetry; percentage and wavelength 
dependence of LP). The three classes of solution are:

\begin{itemize}
\item CP produced mainly by dichroic extinction.

\item CP produced mainly by dichroic scattering.

\item CP having significant contributions from both mechanisms.
\end{itemize}

With such a diversity of solutions it is not possible to determine which mechanism 
has produced the observed CP. We note however, that grain axis ratios of only 1.5:1
(and perhaps less) are sufficient to produce successful solutions by dichroic extinction,
whereas axis ratios approaching 3:1 are needed for solutions in which dichroic
scattering dominates (see section 3.1). Previous estimates have allowed axis ratios in 
the range from 1.1 to 3 in the BN-KL region (eg. Hildebrand \& Dragovan 1995), so a 3:1 
ratio cannot be ruled out. Nevertheless, the smaller grain ratios associated with 
dichroic extinction solutions are more consistent with the typical values derived from
linear polarimetry of star formation regions.

With regard to production of UV CP, a grain axis ratio less than the value of
3 used earlier in this work would not necessarily reduce the degree of polarisation. The 
successful scenario 2(a), and others dependent on dichroic extinction, can compensate 
for a smaller axis ratio by using an increased optical depth. Inspection of the 
extinction matrix elements indicates that if the SACS1 mix were used with an axis 
ratio of 2, the optical depth would have to be increased by a factor of only 1.56, 
(eg. from $\tau=3$ per slab of cloud to $\tau=4.7$) in order to produce the same CP. 
However, for 
an axis ratio of 1.5, the optical depth would have to be increased by a factor of 2.70
(corresponding to $\tau=8.1$ per slab), which would reduce the emergent flux even from
very luminous O-type stars to the point where other light sources would probably 
dominate the local radiation field, at least in crowded star forming regions like Orion.
Thus, an axis ratio of 2 is no obstacle to production of $\sim 10\%$ net UV CP, but
with an axis ratio of 1.5 or less the UV CP is unlikely to exceed 5 or 6\%. 

\normalsize
\section{Discussion}

We have explored a number of scenarios which attempt to produce high CP
in star formation regions via dichroic scattering and/or dichroic extinction. 
Scenarios which rely principally upon dichroic extinction have produced 
UV net CP$\approx10\%$. This is lower than the CP produced in infrared models
owing to the much lower birefringence in the UV for plausible grain size distributions
(see Figure 2). The unscattered component of the radiation field can have net 
CP $\ge 20\%$ in the UV (see Table 1 and Table 2) but the scattered component
reduces the overall CP in the scenarios considered here. Scenarios which 
rely principally on dichroic scattering rather than dichroic extinction are much less 
successful, due to the forward throwing ultraviolet phase function. 
A scenario not considered here is that a low mass YSO or T Tauri star might produce 
high ultraviolet CP in parts of its own envelope. Computation of this mode by 
Wolff et al.(2000) with spherical grains yielded very low CP due to the symmetry of the 
system but computation with aligned non spherical grains has yet to be explored.
It should be noted that this work does not exhaust all the possibilities for 
production of high UV CP, so we do not attempt to put an upper limit on the CP which 
may occur in star formation regions.

The production of 10\% UV CP is still sufficient to be encouraging for the hypothesis 
that CP in star forming regions may produce an initial enantiomeric excess in prebiotic
molecules by asymmetric photolysis.
A CP of 10\% can be produced by dichroic extinction even if the grain axis ratio
is only 2:1 (see Section 4). Furthermore, in Scenarios 2(a) and 2(b) (which involve
dichroic extinction by one or more interstellar cloud sheets) the CP of the radiation 
field in locations well beyond the final cloud sheet does not vary strongly with 
spatial position. This is an attractive feature for crowded star forming regions 
such as Orion, where YSOs are in constant motion. (Close encounters between YSOs 
on 10$^4$~AU scales are fairly common in Orion (Scally \& Clarke 2001) and last 
$\sim 10^4$~yr for typical relative velocities of 3 kms$^{-1}$). By contrast, in the 
scenarios considered in section 3.2, in which the CP is produced in the envelope of a 
high mass YSO, a passing proto-sun would often experience CP which varies significantly 
in space, and in many cases the incident LCP and RCP integrated over the path would 
average to near zero.
 
The efficiency of asymmetric photolysis
as a function of CP percentage is not well quantified at present. Experiments
by Bonner \& Dean (2000) show that elliptically polarised light produces a smaller
enantiomeric excess than 100\% circularly polarised light. Their optical system with the 
principal axis of a quarter waveplate inclined at 22$^{\circ}$.5 to the polarisation 
plane of a 100\% linearly polarised beam should theoretically have produced 70.7\% CP and 
70.7\% LP (the Stokes vector elements add in quadrature to 100\%). The measured 
enantiomeric excess produced in Leucine with this set-up was 73-83\% of that produced by 
100\% circularly polarised light, which tends to support a roughly linear 
relationship with CP, defined as CP=$|V/I|$. Further experiments of this type would be 
very useful. 

We note that circular dichroism has not yet been measured for the amino acids found in 
meteorites (see Bailey 2001) but experiments with other amino acids (Flores et al., 1977),
albeit dissolved in solution, suggest that CP well in excess of 10\% is necessary to 
produce the large enantiomeric excesses measured in meteorites. Even 99\% asymmetric 
photolysis by radiation with 10\% CP would produce an excess which is below 1\%, unless 
the enantiomeric excess vs. CP relationship is highly non linear. It is possible that 
asymmetric photolysis in star forming regions acts first on molecules with higher 
circular dichroism, and the resulting enantimeric excess is then transferred chemically 
to other chiral molecules. For example, some of the 'amino acids' in the Murchison 
meteorite are present only as precursor molecules with unknown circular dichroism. They 
become amino acids only upon extraction by acid
hydrolysis. However, the enantiomeric excess of up to 15\% in isovaline measured in the 
Murchson meteorite remains problematic. Terrestrial homochirality is easier to explain at 
least, since experiments by Shibata et al.(1998) have shown that a small initial 
enantiomeric excess can be greatly amplified by asymmetric autocatalysis.

CP attributed to the dichroic extinction 
mechanism has been observed towards many astrophysical sources, as discussed
in Section 1. However, it is important to increase the size
of the observational database, in order to search for clear examples of CP 
production by dichroic extinction in star formation regions. Observation at shorter 
wavelengths ($\le 1.25~\mu$m) would be helpful since significant alignment of the LP
vectors would then be expected in regions of high CP, owing to the increase in the
K34/K12 ratio (see Figure 2). Any alignment caused by dichroic scattering would be in the
perpendicular direction.
The CP produced by twisting magnetic fields in the diffuse interstellar medium along 
lines of sight toward main sequence stars is $\ll 1\%$, eg. Kemp \& Wolstencroft (1972). 
However, the much higher densities found in star formation regions are likely to produce 
higher CP. We note that a large scale survey of star formation regions in the southern 
sky is currently planned with the Japanese Infrared Survey Facility in South Africa.

\vspace{4mm}
\renewcommand{\baselinestretch}{1}
\normalsize
\textbf{Acknowledgements}\\

"This work was carried out on the Miracle Supercomputer, at the
HiPerSPACE Computing Centre, UCL, which is funded by the U.K. Particle Physics and 
Astronomy Research Council. We thank Janneke Balk of the Oxford University Biochemistry 
Department, Mike Wolff of the Space Telescope Science Institute, Jeremy Yates of 
University College London and Dave Aitken at the University of Hertfordshire for helpful 
advice and discussions. We particularly thank Jim Collett of UHerts for helping us to 
develop parameterised magnetic field structures. We are also grateful to the referee 
for a helpful report. PWL is supported by a PPARC Advanced Fellowship."\\


\textbf{References}\\

Ageorges, N., Fischer, O., Stecklum, B., Eckart, A., Henning, T.: 1996, 
   The Chamaeleon Infrared Nebula: A Polarization Study with High Angular 
    Resolution, {\it Astrophys. J. Letters} 463, pp. 101-104\\

Bailey, J., Chrysostomou, A., Hough, J.H., Gledhill, T.M., McCall, A.,
   Clark, S., Menard, F., and Tamura, M.: 1998, Circular Polarization in Star-Formation
   Regions: Implications for Biomolecular Homochirality, {\it  Science} 281, 672-674\\

Bailey, J.: 2001, Astronomical Sources of Circularly Polarised Light and the Origin of 
   Homochirality, {\it Orig. Life and Evol. Biosphere} 31, 167-183\\

Barber, P.W. and Hill, S.C.: 1990, {\it Light Scattering by Particles: Computational
   Methods}, World Scientific, New York\\

Bonner, W.A.: 1991, The Origin and Amplification of Biomolecular Chirality,
     {\it Orig. Life and Evol. Biosphere} 21, 59-111\\

Bonner, B.A. and Dean, B.D.: 2000, Asymmetric Photolysis with Elliptically Polarised
  Light, {\it Orig. Life and Evol. Biosphere} 30, 513-517\\

Brack, A.: 1998, {\it The Molecular Origins of Life}, CUP, Cambridge, pp. 1-11\\ 

Brand, P.W.J.L., Moorhouse, A., Burton, M.G., Geballe, T.R., Bird, M., and Wade, R.:
   1988, Ratios of molecular hydrogen line intensities in shocked gas: evidence for
   cooling zones, {\it Astrophys. J. Letters} 334, 103-106\\

Cardelli, J.A., Clayton, G.C., and Mathis, J.S.: 1989, The relationship between 
  infrared, optical, and ultraviolet extinction, {\it Astrophys. J.} 345, 245-256\\

Chiang, E.I., Joung, M.K., Creech-Eakman, M.J., Qi, C., Kessler, J.E.,
  Blake, G.A., and van Dishoeck, E.F.: 2001, Spectral Energy Distributions of Passive T 
  Tauri and Herbig Ae Disks: Grain Mineralogy, Parameter Dependences, and Comparison 
  with Infrared Space Observatory LWS Observations, {\it Astrophys. J.} 547, 1077-1089\\

Chrysostomou, A., Hough, J.H., Burton, M.G., and Tamura, M.: 1994, Twisting magnetic
  fields in the core region of OMC-1, {\it Mon. Not. R. Astron. Soc.} 268, 325-334\\

Chrysostomou, A., Menard, F., Gledhill, T.M., Clark, S., Hough, J.H., McCall, A.,
  Tamura, M.: 1997, Polarimetry of young stellar objects - II. Circular polarization 
  of GSS 30, {\it Mon. Not. R. Astron. Soc.} 285, 750-758\\

Chrysostomou, A., Gledhill, T.M., Menard, F., Hough, J.H., Tamura, M., 
   and Bailey, J.: 2000, Polarimetry of Young Stellar Objects - III. Circular 
   Polarimetry of OMC-1, {\it Mon. Not. R. Astron. Soc.} 312, 103-115\\

Cronin, J.R. and Pizzarello, S.: 1997, Enantiomeric excesses in meteoritic amino acids, 
   {Science} 275, 951-955\\

Draine, B.T.: 1985. Tabulated optical properties of graphite and silicate grains,
   {\it Astrophys. J. Supplement} 57, 587-594\\

Engel, M.H. and Macko, S.A.: 1997, Isotopic evidence for extraterrestrial non-racemic 
    amino acids in the Murchison meteorite, {\it Nature} 389, 265-268\\

Fischer, O., Henning, Th., and Yorke, H.W.: 1994, Simulation of polarization maps. 1: 
    Protostellar envelopes, {\it Astron. Astrophys.} 284, 187-209\\

Flores, J.J., Bonner, W.A. and Massey, G.A.: 1977, Asymmetric Photolysis of (RS)-Leucine
   with Circularly Polarised Ultraviolet Light, {\it J. Am. Chem. Soc} 99, 3622-3625\\

Gledhill, T.M.: 1991, Linear polarization maps of bipolar and cometary nebulae - 
    A polarized source interpretation, {\it Mon. Not. R. Astron. Soc.} 252, 138-150\\

Gledhill, T.M., Chrysostomou, A., and Hough, J.H.: 1996, Linear and circular imaging 
   polarimetry of the Chamaeleon infrared nebula, {\it Mon. Not. R. Astron. Soc.} 
   282, 1418-1436\\

Gledhill, T.M. and McCall, A.: 2000. Circular polarization by scattering from spheroidal
  dust grains, {\it Mon. Not. R. Astron. Soc.} 314, 123-137\\

Hamann, F. and Persson, S.E.: 1992, Emission-line studies of young stars. III - 
   Correlations with the infrared excess, {\it Astrophys. J.} 394, 628-642\\

Hildebrand, R.H., and Dragovan, M.: 1995, The shapes and alignment properties of 
   interstellar dust grains, {\it Astrophys. J.} 450, 663-666\\

Hillenbrand, L.A., Strom, S.E., Vrba, F.J, Keene, J.: 1992, Herbig Ae/Be stars - 
   Intermediate-mass stars surrounded by massive circumstellar accretion disks, 
   {\it Astrophys. J.} 397, 613-643\\

Hough, J.H. and Aitken, D.K.: 2003, Polarimetry in the infrared: what can be learned? 
   {\it J. Quantitative Spectroscopy and Radiative Transfer} 79-80, 733-740\\

Joyce, G.F., Vosser, G.M., van Boeckel, C.A.A., van Boom, J.H., Orgel, L.E., van 
   Westrenen, J.: 1984, Chiral Selection in poly(C)-directed synthesis of oligo(G),
   {\it Nature} 310, 602-604. \\

Kemp, J.C. and Wolstencroft, R.D.: 1972, Interstellar Circular Polarization: Data for 
   Six Stars and the Wavelength Dependence, {\it Astrophys. J. Letters} 176, 115-118 \\

Lazarian, A. Mechanical Alignment of Suprathermal Grains: 1995, In: Roberge, W.G., 
  Whittet, D.C.B., editors. {\it Polarimetry of the Interstellar Medium.} ASP conf. 
  series 97, pp. 425-429.\\

Lazarian, A. Physics of Grain Alignment. In: Franco, J., Terlevich,L., López-Cruz, O., 
  Aretxaga, I.: 2000, editors. {\it Cosmic Evolution and Galaxy Formation: Structure, 
  Interactions, and Feedback.} ASP conf. series 215, pp. 69-78\\

Lonsdale, C.J.,Dyck, H.M., Capps, R.W., Wolstencroft, R.D.: 1980, 
  Near-infrared circular polarization observations of molecular cloud sources,
  {\it Astrophys. J. Letters} 238, 31-34\\

Lucas, P.W. and Roche, P.F.: 1998, Imaging Polarimetry of Class I young stellar objects,
    {\it Mon. Not. R. Astron. Soc.} 299, 699-722\\

Lucas, P.W.: 2003, Computation of Light Scattering in Young Stellar Objects, 
   {\it J. Quantitative Spectroscopy and Radiative Transfer} 79-80, 921-938\\

Lucas, P.W., Fukagawa, M., Tamura, M., Beckford, A., Itoh, Y., Murakawa, K., Suto, H., 
   and Hayashi, S.S.: 2003, Near IR polarimetry and Magnetic Field Structure in HL Tau 
   (in prep)\\

Kim, S.-H., Martin, P.G. and Hendry, P.D.: 1994. The size distribution of interstellar 
   dust particles as determined from extinction, {\it Astrophys. J.} 422, 164-175\\

Martin, P.G., Illing, R, and Angel, J.R.P.: 1972. Discovery of interstellar circular 
   polarization in the direction of the Crab nebula, 
   {\it Mon. Not. R. Astron. Soc.} 159, 191-201\\  

Martin, P.G.: 1974, Interstellar polarization from a 
   medium with changing grain alignment, {\it Astrophys. J.} 187, 461-472\\

Mathis, J.S., Rumpl, W., Nordsieck, K.H.: 1977, The size distribution of interstellar 
   grains, {\it Astrophys. J.} 217, 425-433\\

Menard, F., Chrysostomou, A., Gledhill, T.M., Hough ,J.H., Bailey, J.:
   2000, In: Lemarchand, G. and Meech, K.(eds.) {\it 'Bioastronomy 99: A New Era in 
   the Search for Life in the Universe'} (San Francisco) ASP conf. series 213, pp. 
   355-358\\

Minchin, N.R., Hough, J.H., McCall, A., Burton, M.G., McCaughrean, M.J.,
   Aspin, C., Bailey, J.A., Axon, D.J, and Sato, S.: 1991a, Near infrared imaging 
   polarimetry of bipolar nebulae - I. The BN-KL region of OMC-1, {\it Mon. Not. R. 
   Astron. Soc.} 248, 715-729\\

Minchin, N.R., Hough, J.H., Burton, M.G., and Yamashita, T.: 1991b, Near-infrared 
   imaging polarimetry of bipolar nebulae. IV - GL 490, GL 2789 and GL 2136,
   {\it Mon. Not. R. Astron. Soc.} 251, 522-528\\

Mishchenko, M.I., Hovenier, J.W., Travis, L.D.: 2000, {\it Light Scattering by 
   Nonspherical Particles: Theory, Measurements, and Applications} Academic Press, 
   San Diego, 2000.\\

O'Dell, C.R., and Wen, Z.: 1994, Postrefurbishment mission Hubble Space Telescope 
   images of the core of the Orion Nebula: Proplyds, Herbig-Haro objects, and 
   measurements of a circumstellar disk, {\it Astrophys. J.} 436, 194-232\\

Pizzarello, S. and Cronin, J.R.: 2000, {\it Geochim. Cosmochim. Acta} 64, 329-338\\

Pollack, J.B., Hollenbach, D., Beckwith, S., Simonelli, D.P., Roush, T., Fong, W.: 
     1994, Composition and radiative properties of grains in molecular clouds and
     accretion disks. {\it Astrophys. J.} 421, 615-639\\

Preibisch, Th., Ossenkopf, V., Yorke, H.W., and Henning, Th.: 1993, The Influence of 
   ice-coated grains on protostellar spectra, {\it Astron. Astrophys.} 279, 577-588\\

Purcell, E.M.: 1979, Suprathermal rotation of interstellar grains, 
   {\it Astrophys. J.} 231, 404-416\\      

Roberge, W.G.: 1995, Grain Alignment in Molecular Clouds. In: Roberge, W.G., 
   Whittet, D.C.B., editors. {\it Polarimetry of the Interstellar Medium.} ASP conf. 
   series 97, pp. 401-418.\\

Roberts, J.A.: 1984, Supernovae and Life, {\it Nature} 308, 318\\

Rubenstein, E., Bonner, W.A., Noyes, H.P., Brown, G.S.: 1983, Supernovae and Life,
   {\it Nature} 306, 118\\

Scally, A. and Clarke, C.: 2001, Destruction of protoplanetary discs in the Orion Nebula
   Cluster, {\it Mon. Not. R. Astron. Soc.} 325, 449-456\\

Shafter, A. and Jura, M.: 1980, Circular polarization from scattering by 
   circumstellar grains, {\it Astronom. J.} 85, 1513-1519\\

Shakura, N.I., and Sunyaev, R.A.: 1973, Black holes in binary systems. Observational 
   appearance, {\it Astron. Astrophys.} 24, 337-355\\ 

Shibata, T., Yamamoto, J., Matsumoto, N., Yonekubo, S., Osanai, S., and Soai, K.: 1998, 
   Amplification of a Slight Enantiomeric Imbalance in Molecules based on Asymmetric 
   Autocatalysis: the first correlation between high enantiomeric enrichment in a
   chiral molecules and circularly polarised light, {\it J. Am. Chem. Soc.} 120, 
   12157-12158\\

Smith, C.H., Wright, C.M., Aitken, D.K., Roche, P.F., and Hough, J.H.: 2000, 
   Studies in mid-infrared spectropolarimetry - II. An atlas of spectra,
   {\it Mon. Not. R. Astron. Soc.} 312, 327-361\\

Sorrell, W.H.: 1990, The 2175-A feature from irradiated graphitic particles,
   {\it Mon. Not. R. Astron. Soc.} 243, 570-587\\

Stapelfeldt, K.R.; Watson, A.M., Krist, J.E., Burrows, C.J., Crisp, D., 
   Ballester, G.E., Clarke, J.T., Evans, R.W., Gallagher, J.S.III, Griffiths, R.E.,
   Hester, J.J., Hoessel, J.G., Holtzman, J.A., Mould, J.R., Scowen, P.A., 
   Trauger, J.T.: 1999, A Variable Asymmetry in the Circumstellar Disk of HH 30,
     {\it Astrophys. J. Letters} 516, 95-98\\

Tamura, M., Hough, J.H., Hayashi, S.S.: 1995, 1 Millimeter Polarimetry of Young Stellar 
  Objects: Low-Mass Protostars and T Tauri Stars, {\it Astrophys. J.} 448, 346-355\\

Ulrich, R.K.: 1976, An infall model for the T Tauri phenomenon, {\it Astrophys. J.} 
    210, 377-391\\

Whitney, B.A., and Wolff, M.: 2002, Scattering and Absorption by Aligned Grains in 
   Circumstellar Environments, {\it Astrophys. J.} 574, 205-231\\

Whittet, D.C.B.: 1992, {\it Dust in the Galactic Environment}, pub IOP, London, 
   pp. 79-104\\

Wright, C.M.: 1994, {\it Mid-Infrared Spectropolarimetry of Molecular Cloud Sources:
   Magnetic Fields and Dust Properties} PhD thesis, Univ. College, Univ. of New
   South Wales, Sydney\\

Wolff, M.J., Whitney, B.A., Clayton, G.C., Ferris, J.P., Whittet, D.C.B. and Sofia, 
  U.J.: 2000, Circular Polarization in Protostellar and pre-Main Sequence 
  Environments:  Implications for Enantiomeric Excesses in Amino Acids, in proc. 
  {\it First Astrobiology Science Conference} NASA/Ames Research Center, pp. 116-119\\

Wood, K. and Reynolds, R.J.: 1999, A Model for the Scattered Light Contribution and 
   Polarization  of the Diffuse H$\alpha$ Galactic Background, {\it Astrophys. J.} 
   525, 799-807\\

Wood, K., Wolk, S.J., Stane, K.Z., Leussis, G., Stassun, K., Wolff, M., and Whitney, B.:
   2000, Optical Variability of the T Tauri star HH30 IRS, 
   {\it Astrophys. J. Lett.} 542, L21-24\\

\end{document}